# Third Generation MEEG Source Connectivity Analysis Toolbox (BC-VARETA 1.0) and Validation Benchmark

Eduardo Gonzalez-Moreira[1,3#], Deirel Paz-Linares[1,2#], Ariosky Areces-Gonzalez[1,4], Rigel Wang[1] and Pedro A. Valdés-Sosa[1,2*]

**# *contributed equally as first authors***

eduardo.g.m@neuroinformatics-collaboratory.org

deirel.paz@neuroinformatics-collaboratory.org

*** *corresponding author***

pedro.valdes@neuroinformatics-collaboratory.org

Affiliations:

*(1) The Clinical Hospital of Chengdu Brain Science Institute, MOE Key Lab for Neuroinformation, University of Electronic Science and Technology of China, Chengdu, China*
*(2) Cuban Neuroscience Center, La Habana, Cuba*
*(3) Centro de Investigaciones de la Informática, Universidad Central "Marta Abreu" de Las Villas*
*(4) Departamento de Informatica, Universidad de Pinar del Rio, Pinar del Rio, Cuba.*

## Summary

This paper presents a new toolbox for MEEG source activity and connectivity estimation: "Brain Connectivity Variable Resolution Tomographic Analysis version 1.0" (BC-VARETA 1.0). It relies on the third generation of nonlinear methods for MEEG Time Series analysis. Into the state of the art of MEEG analysis, the methodology underlying our tool (BC-VARETA) brings out several assets. First: Constitutes a truly Bayesian Identification approach of Linear Dynamical Systems in the Frequency Domain, grounded in more consistent models (third generation) for the joint nonlinear estimation of MEEG Sources Activity and Connectivity. Second: Achieves Super-Resolution, through the iterative solution of a Sparse Hermitian Sources Graphical Model that underlies the Connectivity Target Function. Third: Tackles efficiently in High Dimensional and Complex set up the estimation of connectivity, those constituting technical issues that challenge current MEEG source analysis methods. Fourth: Incorporates priors at the connectivity level by penalizing the groups of variables, corresponding to the Gray Matter anatomical segmentation, and including a probability mask of the anatomically plausible connections, given by synaptic transmission in the short-range (spatially invariant empirical Kernel of the connections strength decay with distance) and long-range (White Matter tracks connectivity strength from Diffusion Tensor Imaging). Along with the implementation of our method, we include in this toolbox a benchmark for the validation of MEEG source analysis methods, that would serve for the evaluation of sophisticated methodologies (third generation). It incorporates two elements. First: A realistic simulation framework, for the generation of MEEG synthetic data, given an underlying source connectivity structure. Second: Sensitive quality measures that allow for a reliable evaluation of the source activity and connectivity reconstruction performance, based on the Spatial Dispersion and Earth Movers' Distance, in both source and connectivity space.

# 1 INTRODUCTION

Currently there is a consent in the neuroscience community that neural communication patterns between brain regions (brain networks) play a crucial role in brain function at behavior and cognition levels (*Avena-Koenigsberger, 2018*). In the neuroscience field, brain networks topology builds on dense synaptic connections, between individual neurons at the microscale, which are modeled at the mesoscale as nodes with certain connection patterns: directed or undirected (*Salvador, et al., 2005, Estrada, 2012*). This has been evidenced by the progress of invasive or noninvasive brain imaging techniques up until now.

Histological studies in the past uncovered several aspects of brain networks organization at the mesoscale (*Brodmann, 1909; Hässler, 1967; Passingham, 1973; Allman, 1988; Collins et al., 2005; Zilles et al., 1979, 2004; Bailey and Bonin, 1950; Economo and Koskinas, 1925; Jones, 1962*). First: The brain cortex possesses a columnar organization with seven layers (granular or agranular) of morphologically and functionally different neuron types (pyramidal, spiny and smooth). Second: The synaptic connections follow a layer specific organization. Spiny striate cells in granular layers act as receptors of excitatory impulses from pyramidal neurons in the infra-agranular layers or inhibitory also from pyramidal neurons in supra-agranular layers. This takes effect in inter-layer communication patterns of three types: intracolumnar (directed connections) and intercolumnar in both short range (lateral undirected connections) and long range (forward or backward directed connections).

The analysis of Blood Oxygenation Dependent (BOLD) signal registered by Functional Magnetic Resonance Imaging (fMRI), can reflect the neural correlates (responses and connectivity) of brain function during task or resting state. Also based on MRI, the Diffusion Weighted Imaging (DWI) allows to extract the probabilistic maps of the long-range connectivity due to white matter tracks. These constitute at most, what is available to investigate the connectivity noninvasively, with spatial resolution that can reach the columnar level, providing reliable correlates of spatially distributed neural activity. Unfortunately, these techniques do not reflect directly the neural dynamics or synaptic transmission. The BOLD signal is a consequence of a slow (spans over seconds) metabolic/hemodynamic cascade which is activated by synaptic activity, thus it does not reach the milliseconds time-scale of faster brain rhythms. DWI provides structural probabilistic maps of the plausible connections, based on the diffusion of water across White Matter tracks, but cannot reveal precisely the pathways that take effect in neural communication.

Non-invasive electrophysiological recordings, such as magneto/electroencephalography (MEEG), bring an ideal scenario to cover the gap of other slower and indirect imaging methods, e.g. the previously cited fMRI. Its direct link to local field potentials (associated to synaptic events) and high temporal resolution (milliseconds) allows the tracking of the neural processes underlying human perception and cognition (*Schomer and Lopes da Silva 2011*, *Hämäläinen, et al., 1993*). This is explained given the local field potential of synchronized neural activity within neural masses (generators), which creates a noninvasively observable Primary Current Density (PCD). An accurate estimation of the PCD given these signals would thus provide a representation of the neural dynamic, therefore the MEEG based connectivity analysis constitutes a strong approach to study brain functional networks in Resting State (RS) or Event Related Potential (ERP) (*Schoffelen and Gross, 2009; Smit, et al., 2008*).

Unfortunately, the analysis of MEEG sources activity (PCD reconstruction) constitutes a severely ill-posed problem, i.e. the MEEG Inverse Problem (MEEG-IP). The reasons for this are two: First, the small amount of data (hundreds of scalp recording points) compared to the large amount of Gray Matter generators (thousands) to be estimated. Second, smearing of the sources activity when projected from the generators space to the scalp sensors via the Lead Field, this has been pinpointed "Volume Conduction Effect". Due to the latter the number of sensors that would carry relevant information about source activity is limited, rendering the former situation a theoretical (not practical) shortcoming (*Hassan and Wendling, 2018*). It is affirmed that the only solution to this problem would be developing MEEG source analysis methods which might be flexible enough to incorporate priors on the spatio-temporal patterns source activity, or what would be better: jointly on source activity and connectivity.

Up to now, the MEEG source analysis models can be classified into the Bayesian formalism as three main generations, accounting for how these models make use of prior information about connectivity (covariances or precisions). First Generation: Uses a fixed covariance structure while solving the source activity estimation by a linear formula, e.g. Minimum Norm (MN) (*Hämäläinen and Ilmoniemi, 1994*) and LORETA (*Pascual-Marqui et al., 1994*). Second Generation: Regards embedded priors on the source activity and a diagonal covariance (variances) structure, while the estimation is tackled by nonlinearly dependent formulas of both source activity and connectivity, e.g. Exact LORETA (eLORETA) (*Pascual-Marqui, 2002*), Multiple Penalized Least Squares (MPLS) (*Vega-Hernández et al., 2008*) and Structured Sparse Bayesian Learning (SSBL) (*Paz-Linares et al., 2017*). Third Generation: It does as the second generation but with a full covariance structure, e.g. Variable Resolution Tomographic Analysis (VARETA) (*Valdes-Sosa, 1996, Bosch-Bayard, et al., 2001*) and Restricted Likelihood Maximization (ReLM) (*Patterson and Thompson, 1971, Harville, 1977; Friston et al., 2007; Wipf et al., 2009; Belardinelli et al., 2012; Wu et al., 2016*).

Nevertheless, ESI methods have been developed mainly to estimate activation and not connectivity. Indistinctively, these have been implemented as a first stage before connectivity postprocessing (second step) or statistical analysis of the Sources' time series (*Sakkalis, 2011*, *Bastos and Schoffelen, 2016*), such as Granger Causality (*Granger, 1969*), Dynamical Causal Models (DCM) (*Penny, 2004*), frequency domain connectivity measures like Coherence (Coh) (*Tucker, et al. 1986; Srinivasan, et al., 2007*; *Guillon, et al., 2017*), Partial Coherence (PCoh) (*Lopes da Silva, et al., 1980*), Directed Coherence (DC) and Partial Directed Coherence (PDC) (*Baccalá and Sameshima, 2001*), population statistical analysis of the results for source activity and connectivity features extraction (*Hipp et al., 2012*; *Babiloni et al., 2005*; *Brookes, 2001*).

The use of first and second generation MEEG methods is quite generalized into the state of the art, meanwhile, using third generation methods seems limited only to theoretical studies. From this, a severe methodological error stands out: the use of the "two steps" approach towards connectivity analysis, which renders the estimation unprecise due the ill-conditioning of the MEEG-IP. This conceptual problem relies on the idea that the simultaneously estimation of activation and connectivity has been unappreciated. This is something totally deliberated given the state-space nature of the MEEG model (*Galka, et al., 2004, Valdes-Sosa, 2004*), and its subsequent interpretation as Gaussian Graphical Models.

This work serves as continuation to theoretical developments of a model meant to revendicate the third generation of MEEG source analysis methods: Brain Connectivity Variable Resolution Tomographic Analysis (BC-VARETA) (*Paz-Linares&Gonzalez-Moreira et al., 2018a; Paz-Linares&Gonzalez-Moreira et al., 2018b*). This was strongly motivated by the idea on the unification of a well stablished third generation method VARETA (*Valdes-Sosa, 1996*, *Bosch-Bayard, et al., 2001*) and the theory of high dimension covariance of precision matrices (*Maurya, 2016*, *McGillivray, 2016*, *Ledoit and Wolf 2015*, *Cai, et al., 2016*, *Adegoke, et al., 2018*).

## Objectives

In this paper, we present a third generation opensource toolbox based on BC-VARETA model (BC-VARETA 1.0), for super-resolution and high dimensional MEEG connectivity analysis. This will be possible due to the implementation of an efficient algorithm for the Group LASSO (structured sparsity) model directly on the source precision matrix. It allows for incorporating in the estimation procedure information about brain anatomical areas or prior probability maps of the connectivity in the short and long range. This approach is meant for the analysis stationary time series in the frequency domain, through the estimation of an underlaying Hermitian Embedded Gaussian Graphical Models (HEGGM), that arises from the Bayesian representation of Linear State Space Models (LSSM). We built a validation Benchmark based on in simulations, which incorporates inverse crime evaluation, noise from biological and instrumentation origin, realistic sources set up and quality measures of both source activity and connectivity reconstruction. This Benchmark sets the conditions for further evaluation of third generation MEEG methods. Finally, we present a study devoted demonstrate the efficacy of our source analysis tool in both synthetic and real examples.

The technical route of BC-VARETA 1.0 consists in the following steps:

1- Lead Field computation by the extraction od head model from the individual subject T1 MRI.
2- Definition of anatomical regions on the individual subject cortical surface.
3- Extraction data samples by the Discrete Fourier Transform (DFT) of the sensors time series in multiple windows.
4- Initial screening of sources by the second generation method SSBL, which allows to reduce the source space dimensionality.
5- Precision matrix estimation by BC-VARETA method.
6- Computation of the Partial Coherence as measure of the Neural generators Functional Connectivity.

# 2 METHODOLOGY

## 2.1 Inference with BC-VARETA model

### 2.1.1 Joint Bayesian Model of Source Activity and Connectivity

Within BC-VARETA framework, the frequency domain generative model of the MEEG signal is expressed mathematically by a hierarchically conditional Gaussian Graphical Models, comprising the Bayesian representation of two components[1]: 1-Forward Model (Observation Equation) as Data Likelihood. 2-Linear Dynamical Model (State Equation) as Source Prior. See in Figure 1 the More-Penrose diagram corresponding to these equations.

$$\boldsymbol{v}_m(\omega)|\boldsymbol{\iota}_m(\omega),\sigma_\xi^2(\omega) \backsim N_{\mathrm{p}}^{\mathbb{C}}\left(\boldsymbol{v}_m(\omega)\middle|\mathbf{L}_{\boldsymbol{v\iota}}\boldsymbol{\iota}_m(\omega),\sigma_\xi^2(\omega)\mathbf{R}_{\xi\xi}\right); m \in \mathbb{M}; \omega \in \mathbb{F} \quad [2.1.1]$$

$$\boldsymbol{\iota}_m(\omega)|\boldsymbol{\Theta}_{\boldsymbol{\iota}}(\omega) \backsim N_{\mathrm{q}}^{\mathbb{C}}\left(\boldsymbol{\iota}_m(\omega)\middle|\mathbf{0},\boldsymbol{\Theta}_{\boldsymbol{\iota}}^{-1}(\omega)\right); m \in \mathbb{M}; \omega \in \mathbb{F} \quad [2.1.2]$$

The Data $\left(\boldsymbol{v}_m(\omega)\right)_{\mathrm{p}\times 1}$, in the sensor space $\mathbb{E}$, and Parameters $\left(\boldsymbol{\iota}_m(\omega)\right)_{\mathrm{q}\times 1}$, in the cortical manifold space $\mathbb{G}$, are complex vectors (Fourier Coefficients) from the representation in the frequency domain $\mathbb{F}$ (at a single spectral component $\omega \in \mathbb{F}$) of the MEEG signal and sources activity. The subscript p represents the number of sensors, q the number of generators and m the number of time windows in which the Fourier coefficients are computed. $N^{\mathbb{C}}$ represents the Circularly Symmetric Complex Gaussian distribution. The Source to Data Transfer Operator (SDTO) $(\mathbf{L}_{\boldsymbol{v\iota}})_{\mathrm{p}\times\mathrm{q}}$ is obtained from the discretization of a specific head model Lead Field. The Data conditional covariance of [2.1.1] represents that of a noisy process $\boldsymbol{\xi}$ at the sensors, which is assumed to be composed by two factors. First: A frequency dependent nuisance variance $\sigma_\xi^2(\omega)$ (Hyperparameter). Second: An ad-hoc symmetric positive definite matrix $\left(\mathbf{R}_{\xi\xi}\right)_{\mathrm{p}\times\mathrm{p}}$ of the sensors correlation structure, that can be determined experimentally. The frequency dependent Source Precisions (SP) matrix $\left(\boldsymbol{\Theta}_{\boldsymbol{\iota}}(\omega)\right)_{\mathbf{q}\times\mathbf{q}}$ of [2.1.2] (Hyperparameter) represents the undirected connectivity.

---

[1] *Uppercase and lowercase bold scripts denote matrices and vectors respectively, the observable quantities are denoted by Latin scripts while the unobserved by Greek scripts. The scalars are denoted by lowercase scripts.*

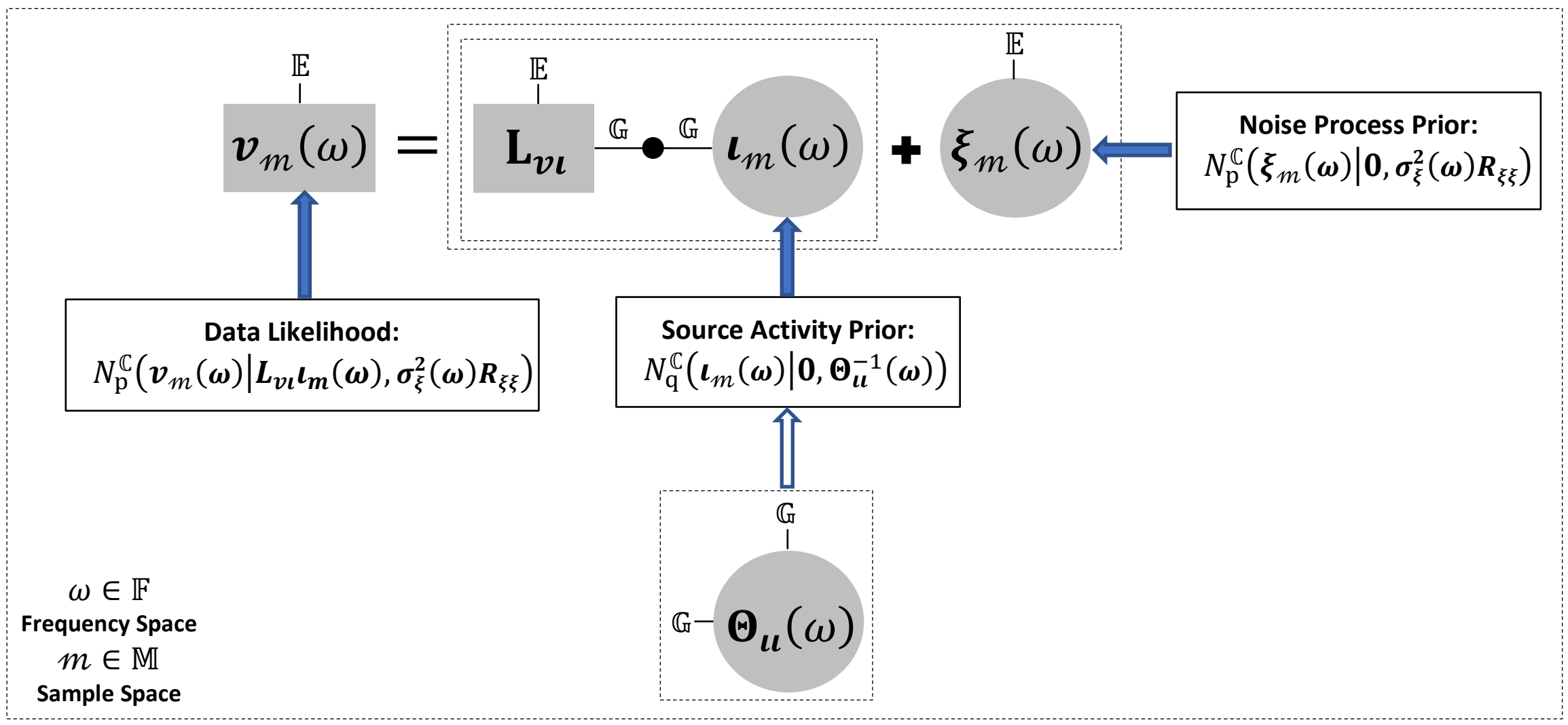

***Figure 1:*** *More-Penrose diagram of the MEEG Source Activity and Connectivity Bayesian Model. The model Variables are represented with gray circles and the priors with white rectangles. The filled arrows represent Random Variables generation by a specific pdf and the unfilled arrows the corresponding pdf parametrization.*

### 2.1.2 Specification of Hyperparameters Priors

BC-VARETA specifies priors on the Hyperparameters space, i.e. mathematically denoted $\Xi(\omega)$. The use of a Jeffrey Improper distribution (*Jeffrey, 1946*), as for nuisance variance prior, informs the model about a noise inferior threshold $\epsilon$ that can be determined experimentally. This improves the convexity of the whole estimation and thus the numerical control.

$$\sigma_\xi^2(\omega) \backsim Exp\left(1/\sigma_\xi^2(\omega)\,\middle|\epsilon\mathrm{mp}\right) \qquad [2.1.3]$$

For the source precision matrix prior some options among the hierarchical complex LASSO (mixtures of Normal and Exponential ***pdf***) family are considered into the theoretical formulation, particularly the simples matrix L1 norm was used in previous validations. Nevertheless, the matrix Group LASSO constitutes the best alternative to our interest incorporate two essential types of priors. First: Brain anatomical regions which define the group penalization of the precision matrix elements, i.e. $\mathbb{G}_i \subset \mathbb{G}$ of size $\mathrm{q}_i$ where $i$ counts the regions $i = 1 \cdots \mathbb{q}$. Second: Prior Connectivity Maps (PCM) in the short long range $\mathbf{A}$. That is defined for the short range as a deterministic spatially invariant empirical Kernel of the connections strength decay with distance. For the long-range connections, it is given by probabilistic maps of the White Matter tracks connectivity strength from Diffusion Tensor Imaging (DTI). Bellow we present the priors of the "Connectivity" $\mathbf{\Theta}_{\boldsymbol{u}}(\omega)$, "Elementwise Connectivity Variances" $\mathbf{\Gamma}(\omega)$ and "Anatomical Areas Connectivity Selector" $\mathbf{\Upsilon}(\omega)$ which take part on the model estimation, see Figure 2 right branch.

$$\mathbf{\Theta}_{\boldsymbol{u}}(\omega) \backsim \prod_{ij=1}^{\mathrm{q}} N_1\left(\left|\left(\mathbf{\Theta}_{\boldsymbol{u}}(\omega)\right)_{ij}\right|\middle|0, \frac{\Gamma_{ij}^2(\omega)}{\mathrm{m}\lambda(\omega)}\right) \quad [2.1.4]$$

$$\Gamma_{ij}^2(\omega) = \frac{\Upsilon_{ij}^2(\omega)}{\mathrm{m}\lambda(\omega)\mathrm{A}_{ij}^2}; (i,j) \in \mathbb{G}_i \times \mathbb{G}_j; ij = 1 \cdots \mathrm{q} \quad [2.1.5]$$

$$\mathbf{\Upsilon}^2(\omega) \backsim \prod_{ij=1}^{\mathrm{q}} Exp\left(\Upsilon_{ij}^2(\omega)\middle|\frac{1}{2}\right) \quad [2.1.6]$$

The inference of the HEGGM in the whole space of generators is theoretically impossible due to nonactive areas that foil the positive definiteness of the precision matrix. To avoid this situation, we initialize BC-VARETA by preselecting those brain anatomical regions that might be active for the subsequent SP estimation. The initial screening of potential brain generators is done by the SSBL algorithm for the Group LASSO model, in similar fashion as formulas [2.1.5] and [2.1.6] but at the source level (Second Generation method). It is formulated through priors on a reparameterization that constraints the covariance matrix to the diagonal "Variances" $\mathbf{\Theta}_{\boldsymbol{u}}(\omega) = diag\left(\boldsymbol{\sigma}_{\boldsymbol{\iota}}^2(\omega)\right)^{-1}$, which are controlled by an "Anatomical Areas Selector" $\boldsymbol{\gamma}(\omega)$, see Figure 2 left branch.

$$\left(\boldsymbol{\sigma}_{\boldsymbol{\iota}}^2(\omega)\right)_i = \frac{\gamma_i^2(\omega)}{\mathrm{m}\lambda(\omega)}; i \in \mathbb{G}_i; i = 1 \cdots \mathrm{q} \quad [2.1.7]$$

$$\boldsymbol{\gamma}^2(\omega) \backsim \prod_{i=1}^{\mathrm{q}} Exp\left(\gamma_i^2(\omega)\middle|\frac{1}{2}\right) \quad [2.1.8]$$

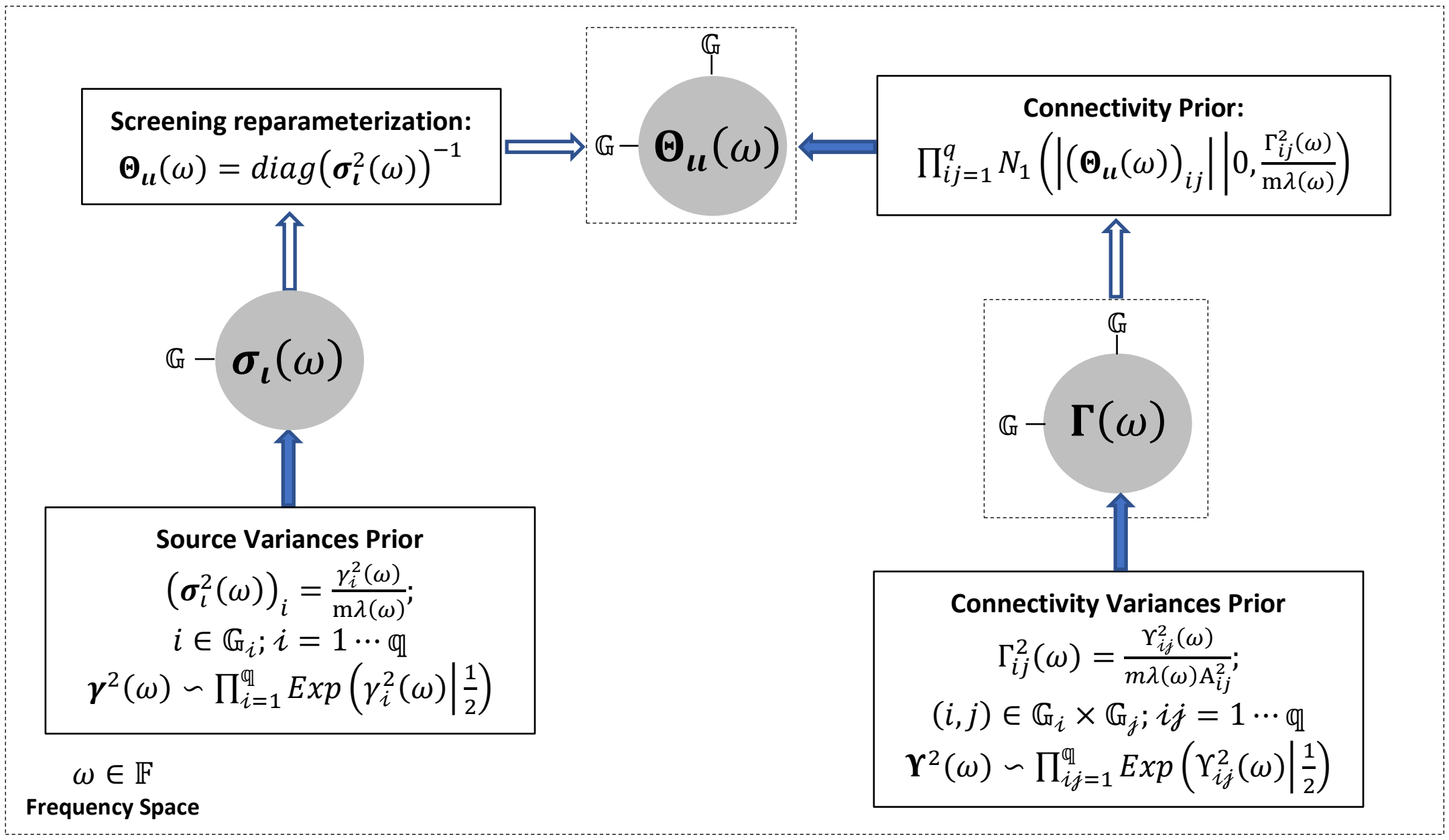

***Figure 2:*** *More-Penrose diagram of the Precision Matrix Priors of Screening (left branch) and Joint Source Activity and Connectivity model (right branch). The model Variables are represented with gray circles and the priors with white rectangles. The filled arrows represent Random Variables generation by a specific pdf and the unfilled arrows the corresponding pdf parametrization.*

### 2.1.3 Expectation Maximization for Source Activity and Connectivity Analysis

BC-VARETA tackles explicit Parameters and Hyperparameters Maximum Posterior Analysis (MAP), independently for each frequency component, via the Expectation-Maximization (EM) algorithm (*McLachlan, 2007*). It consists of two steps that are embedded into an iterative scheme.

**First:** Compute the Data Expected Type II Log-Likelihood $Q\left(\Xi(\omega),\widehat{\Xi}^{(k)}(\omega)\right)$, given for the $k$-th iteration Hyperparameters, i.e. "Expectation" of Data and Parameters joint ***pdf*** for all samples $p\left(\{\boldsymbol{v}_m(\omega),\boldsymbol{\iota}_m(\omega)\}_{m=1}^{\mathrm{m}}\middle|\Xi(\omega)\right)$ by the Parameters posterior ***pdf*** $p\left(\{\boldsymbol{\iota}_m(\omega)\}_{m=1}^{\mathrm{m}}\middle|\{\boldsymbol{v}_m(\omega)\}_{m=1}^{\mathrm{m}},\widehat{\Xi}^{(k)}(\omega)\right)$.

**Second:** Re-estimation of the Hyperparameters $\widehat{\Xi}^{(k+1)}(\omega)$, i.e. "Maximization" of the Hyperparameters Iterated Posterior ***pdf*** $p\left(\Xi(\omega)\middle|\{\boldsymbol{v}_m(\omega)\}_{m=1}^{\mathrm{m}},\widehat{\Xi}^{(k)}(\omega)\right)$, given by the combination of the Data Expected Type II Likelihood $exp\left(Q\left(\Xi(\omega),\widehat{\Xi}^{(k)}(\omega)\right)\right)$ and Hypermeters Prior ***pdf*** $p(\Xi(\omega))$.

Data Expected Type II Log-Likelihood

$$Q\left(\Xi(\omega),\widehat{\Xi}^{(k)}(\omega)\right)=-\frac{\mathrm{m}}{\sigma_\xi^2(\omega)}tr\left(\left(\mathbf{I}_\mathrm{p}-\mathbf{L}_{\boldsymbol{v\iota}}\breve{\mathbf{T}}_{\boldsymbol{\iota v}}^{(k)}(\omega)\right)^\dagger\mathbf{R}_{\xi\xi}^{-1}\left(\mathbf{I}_\mathrm{p}-\mathbf{L}_{\boldsymbol{v\iota}}\breve{\mathbf{T}}_{\boldsymbol{\iota v}}^{(k)}(\omega)\right)\mathbf{S}_{\boldsymbol{vv}}(\omega)\right)\cdots$$

$$-\mathrm{mp}\,log\left(\sigma_\xi^2(\omega)\right)-\frac{\mathrm{m}}{\sigma_\xi^2(\omega)}tr\left(\mathbf{L}_{\boldsymbol{v\iota}}^{\mathcal{T}}\mathbf{R}_{\xi\xi}^{-1}\mathbf{L}_{\boldsymbol{v\iota}}\breve{\boldsymbol{\Sigma}}_{\boldsymbol{\iota}}^{(k)}(\omega)\right)+\mathrm{m}\,log|\boldsymbol{\Theta}_{\boldsymbol{\iota}}(\omega)|-\mathrm{m}\,tr\left(\boldsymbol{\Theta}_{\boldsymbol{\iota}}(\omega)\breve{\boldsymbol{\Psi}}_{\boldsymbol{\iota}}^{(k)}(\omega)\right)\quad[2.1.9]$$

Formula [2.1.10] depends on the free Hyperparameters explicitly and implicitly, though some auxiliary quantities, on the Hyperparameters estimated in the previous iteration. The matrix $\mathbf{S}_{\boldsymbol{vv}}(\omega)$ represents the Data Empirical Covariance (DEC). The auxiliary quantities are the Data to Sources Transference Operator (DSTO) $\breve{\mathbf{T}}_{\boldsymbol{\iota v}}^{(k)}(\omega)$, Parameters Posterior Covariance (PPC) matrix $\breve{\boldsymbol{\Sigma}}_{\boldsymbol{\iota}}^{(k)}(\omega)$, and Sources Effective Empirical Covariance (ESEC) matrix $\breve{\boldsymbol{\Psi}}_{\boldsymbol{\iota}}^{(k)}(\omega)$. The ESEC is defined as the composition of two quantities: The (SPC) matrix and the Source Empirical Covariance (SEC) matrix estimator $\hat{\mathbf{S}}_{\boldsymbol{\iota}}^{(k)}(\omega)$. Bellow we present the formulas for the computation of each auxiliary quantity and estimator involved in the maximization step.

DEC: $$\mathbf{S}_{\boldsymbol{vv}}(\omega)=\frac{1}{\mathrm{m}}\sum_{m=1}^{\mathrm{m}}\boldsymbol{v}_m(\omega)\boldsymbol{v}_m^\dagger(\omega)\quad[2.1.10]$$

SPC: $$\breve{\boldsymbol{\Sigma}}_{\boldsymbol{\iota}}^{(k)}(\omega)\leftarrow\left(\frac{1}{\hat{\sigma}_\xi^2(\omega)}\mathbf{L}_{\boldsymbol{v\iota}}^{\mathcal{T}}\mathbf{R}_{\xi\xi}^{-1}\mathbf{L}_{\boldsymbol{v\iota}}+\widehat{\boldsymbol{\Theta}}_{\boldsymbol{\iota}}^{(k-1)}(\omega)\right)^{-1}\quad[2.1.11]$$

DSTO: $$\breve{\mathbf{T}}_{\boldsymbol{\iota v}}^{(k)}(\omega)\leftarrow\frac{1}{\hat{\sigma}_\xi^2(\omega)}\breve{\boldsymbol{\Sigma}}_{\boldsymbol{\iota}}^{(k)}(\omega)\mathbf{L}_{\boldsymbol{v\iota}}^{\mathcal{T}}\mathbf{R}_{\xi\xi}^{-1}\quad[2.1.12]$$

SEC: $$\hat{\mathbf{S}}_{\boldsymbol{\iota}}^{(k)}(\omega)\leftarrow\breve{\mathbf{T}}_{\boldsymbol{\iota v}}^{(k)}(\omega)\mathbf{S}_{\boldsymbol{vv}}(\omega)\breve{\mathbf{T}}_{\boldsymbol{\iota v}}^{(k)\dagger}(\omega)\quad[2.1.13]$$

ESEC: $$\breve{\boldsymbol{\Psi}}_{\boldsymbol{\iota}}^{(k)}(\omega)\leftarrow\breve{\boldsymbol{\Sigma}}_{\boldsymbol{\iota}}^{(k)}(\omega)+\hat{\mathbf{S}}_{\boldsymbol{\iota}}^{(k)}(\omega)\quad[2.1.14]$$

## 2.2 Implementation, Validation Benchmark and Technical Route of BC-VARETA Toolbox

### 2.2.1 Estimation Formulas and Implementation

The EM strategy of BC-VARETA leads to a compact and explicit estimation procedure. The source activity is represented by the iterated estimator of the ESEC $\check{\mathbf{S}}_{\boldsymbol{\iota}}^{(k)}(\omega)$ given in formula [2.1.13], for both stages of BC-VARETA.

**First:** "Screening" for active anatomical areas, with estimation formulas resulting from Priors [2.1.3], [2.1.7] and [2.1.8] merged to the EM scheme of Section 2.1.3. It defines a reduced cortical space built by the union of those anatomical areas which return nonzero values of the selector $\gamma_{\imath}^2(\omega) \neq 0; \imath = 1 \cdots \mathbb{q}$. The Hyperparameters are set in this case as $\Xi(\omega) = \{\sigma_{\xi}^2(\omega), \boldsymbol{\sigma}_{\iota}^2(\omega), \lambda(\omega)\}$. See Pseudocode 1.

**Second:** "Joint Source Activity and Connectivity Estimation", on the reduced cortical space after screening, with formulas resulting from merging into the EM scheme the Priors [2.1.3], [2.1.4], [2.1.5] and [2.1.6]. Applies Unbiasedness operation to the Connectivity and determine the sparsity patter of nondiagonal elements. Use as Connectivity input to [2.1.11] the ESEC inverse, and then apply to SPC the Unbiased Connectivity sparsity pattern. In this case the Hyperparameters are set as $\Xi(\omega) = \{\sigma_{\xi}^2(\omega), \boldsymbol{\Theta}_{\boldsymbol{\iota}}(\omega), \boldsymbol{\Gamma}(\omega), \lambda(\omega)\}$. See Pseudocode 2.

***Screening Formulas***

$$\check{\gamma}_{\imath}^{2\,(k+1)}(\omega) \leftarrow \mathrm{m}\left(\left(\mathrm{q}_{\imath}^2 + 2\lambda(\omega) \textstyle\sum_{i \in \mathbb{G}_{\imath}} \left(\check{\boldsymbol{\Psi}}_{\boldsymbol{\iota}}^{(k)}(\omega)\right)_{ii}\right)^{\frac{1}{2}} - \mathrm{q}_{\imath}\right); \imath = 1 \cdots \mathbb{q} \qquad [2.1.15]$$

$$\left(\hat{\sigma}_{\boldsymbol{\iota}}^{2\,(k+1)}(\omega)\right)_i \leftarrow \check{\gamma}_{\imath}^{2\,(k+1)}(\omega) / \left(\mathrm{m}\lambda(\omega)\right); i \in \mathbb{G}_{\imath} \qquad [2.1.16]$$

$$\hat{\lambda}^{(k+1)}(\omega) \leftarrow \frac{\mathrm{qm}\hat{\lambda}^{(k)}(\omega)}{2} \textstyle\sum_{i=1}^{\mathrm{q}} \left(\left(\hat{\sigma}_{\boldsymbol{\iota}}^{2\,(k+1)}(\omega)\right)_i \Big/ \left(\check{\boldsymbol{\Psi}}_{\boldsymbol{\iota}}^{(k)}(\omega)\right)_{ii}\right) \qquad [2.1.17]$$

***Pseudocode 1: Screening for active sources***

| | |
|---|---|
| **INPUTS** | $\mathbf{L}_{v\iota}, \mathbf{S}_{vv}(\boldsymbol{\omega}), \mathbf{R}_{\xi\xi}, \{(\mathbb{G}_{\imath}, \mathbf{q}_{\imath}); \imath = \mathbf{1} \cdots \mathbb{q}\}, \mathbf{p}, \mathbf{q}, \mathbf{m}, \boldsymbol{\epsilon}$ |
| **OUTPUTS** | $\hat{\mathbf{S}}_{\boldsymbol{\iota}}(\omega)$, $\hat{\boldsymbol{\Theta}}_{\boldsymbol{\iota}}(\omega)$ |
| **INITIALIZE** | Set: $\hat{\sigma}_{\xi}^{2\,(0)} = 1$, $\hat{\lambda}^{(0)}(\omega) = 1$, $\hat{\boldsymbol{\Theta}}_{\boldsymbol{\iota}}^{(0)}(\omega) = \mathbf{I}_{\mathrm{q}}$ |
| | Compute: "SPC" $\check{\boldsymbol{\Sigma}}_{\boldsymbol{\iota}}^{(0)}(\omega)$ [2.1.11], "DSTO" $\check{\mathbf{T}}_{\boldsymbol{\iota v}}^{(0)}(\omega)$ [2.1.12], "SEC" $\hat{\mathbf{S}}_{\boldsymbol{\iota}}^{(0)}(\omega)$ [2.1.13] |
| | Rescale: "DEC" $\mathbf{S}_{\boldsymbol{vv}}(\omega) \leftarrow \left(\mathrm{q}\, max\left(diag\left(\check{\boldsymbol{\Sigma}}_{\boldsymbol{\iota}}^{(0)}(\omega)\right)\right) / tr\left(\hat{\mathbf{S}}_{\boldsymbol{\iota}}^{(0)}(\omega)\right)\right) \mathbf{S}_{\boldsymbol{vv}}(\omega)$ |
| **OUTER CYCLE** | $k = 1 \cdots$ (counter) |
| ***step 1*** | $\check{\boldsymbol{\Sigma}}_{\boldsymbol{\iota}}^{(k)}(\omega)$ Auxiliary quantity "SPC" [2.1.11] |
| ***step 2*** | $\check{\mathbf{T}}_{\boldsymbol{\iota v}}^{(k)}(\omega)$ Auxiliary quantity "DSTO" [2.1.12] |
| ***step 3*** | $\hat{\mathbf{S}}_{\boldsymbol{\iota}}^{(k)}(\omega)$ Estimator "SEC" [2.1.13] |
| ***step 4*** | $\check{\boldsymbol{\Psi}}_{\boldsymbol{\iota}}^{(k)}(\omega)$ Auxiliary quantity "ESEC" [2.1.14] |
| ***step 5*** | $\check{\boldsymbol{\gamma}}^{(k+1)}(\omega)$ Auxiliary quantity "Anatomical Areas Selector" [2.1.15] |
| ***step 6*** | $\hat{\boldsymbol{\sigma}}_{\boldsymbol{\iota}}^{2\,(k+1)}(\omega)$ Estimator "Source Variances" [2.1.16] |
| ***step 7*** | $\lambda^{(k+1)}(\omega)$ Estimator "Regularization Parameter" [2.1.17] |
| ***step 8*** | $\hat{\sigma}_{\xi}^{2\,(k+1)}(\omega)$ Estimator "Noise Variance" [2.1.23] |
| **END** | |

***Source Precisions Formulas***: $\widehat{\mathbf{\Theta}}_{u}^{(k+1)}(\omega) = lim_{l\to\infty}\left(\widehat{\mathbf{\Theta}}_{u}^{(k+1,l+1)}(\omega)\right)$

$$\breve{\Upsilon}_{\mathcal{ij}}^{(k+1,l)}(\omega) \leftarrow \left(\left(\left(1+4\mathrm{m}\lambda(\omega)\textstyle\sum_{(i,j)\in\mathbb{G}_{\mathcal{i}}\times\mathbb{G}_{\mathcal{j}}}\mathrm{A}_{ij}^{2}\,abs\left(\widehat{\mathbf{\Theta}}_{u}^{(k+1,l)}(\omega)\right)_{ij}^{2}\right)^{\frac{1}{2}}-1\right)^{\frac{1}{2}}\right)\Big/2^{\frac{1}{2}};\ \mathcal{ij}=1\cdots\mathbb{q} \qquad [2.1.18]$$

$$\hat{\Gamma}_{ij}^{(k+1,l)}(\omega) \leftarrow \breve{\Upsilon}_{\mathcal{ij}}^{(k+1,l)}(\omega)\Big/\left(\mathrm{m}^{\frac{1}{2}}\lambda^{\frac{1}{2}}(\omega)\mathrm{A}_{ij}\right);\ (i,j)\in\mathbb{G}_{\mathcal{i}}\times\mathbb{G}_{\mathcal{j}} \qquad [2.1.19]$$

$$\widehat{\mathbf{\Theta}}_{u}^{(k+1,l+1)}(\omega) \leftarrow \hat{\mathbf{\Gamma}}^{(k+1,l)}(\omega)\odot sqrt\left(\left(\left(\breve{\mathbf{\Psi}}_{u}^{(k)^{-1}}(\omega)\oslash\hat{\mathbf{\Gamma}}^{(k+1,l)}(\omega)\right)^{-2}+4\lambda(\omega)\mathbf{I}_{\mathrm{q}}\right)\Big/2\lambda(\omega)\cdots\right.$$

$$\left.-\hat{\mathbf{\Gamma}}^{(k+1,l)}(\omega)\odot\left(\mathbf{\Psi}_{u}^{(k)^{-1}}(\omega)\oslash\hat{\mathbf{\Gamma}}^{(k+1,l)}(\omega)\right)^{-1}\Big/2\lambda(\omega)\right. \qquad [2.1.20]$$

***Unbiased Source Precisions (USP) Formulas*** *(for the condition $\lambda(\omega)=\sqrt{log(q)/m}$)*

$$\left(\widehat{\mathbf{\Theta}}_{u}\right)_{unbiased}^{(k+1)}(\omega) = 2\widehat{\mathbf{\Theta}}_{u}^{(k+1)}(\omega) - \widehat{\mathbf{\Theta}}_{u}^{(k+1)}(\omega)\breve{\mathbf{\Psi}}_{u}^{(k)}(\omega)\widehat{\mathbf{\Theta}}_{u}^{(k+1)}(\omega) \qquad [2.1.21]$$

$$\mathbb{K}_{0}^{(k+1)}(\omega) = \left(\left(\widehat{\mathbf{\Theta}}_{u}\right)_{unbiased}^{(k+1)}(\omega) < diag\left(\widehat{\mathbf{\Theta}}_{u}^{(k+1)}(\omega)\right)diag\left(\widehat{\mathbf{\Theta}}_{u}^{(k+1)}(\omega)\right)^{\mathcal{T}} + \left(\widehat{\mathbf{\Theta}}_{u}^{(k+1)}(\omega)\right)^{.2}\right) \qquad [2.1.22]$$

***Nuisance Variance Formula*** *(common for either of previous stages)*

$$\hat{\sigma}_{\xi}^{2^{(k+1)}}(\omega) \leftarrow \frac{tr\left(\left(\mathbf{I}_{\mathrm{p}}-\mathbf{L}_{v\iota}\breve{\mathbf{T}}_{\iota v}^{(k)}(\omega)\right)^{\dagger}\mathbf{R}_{\xi\xi}^{-1}\left(\mathbf{I}_{\mathrm{p}}-\mathbf{L}_{v\iota}\breve{\mathbf{T}}_{\iota v}^{(k)}(\omega)\right)\mathbf{S}_{vv}(\omega)\right)}{\mathrm{p}} + \frac{tr\left(\mathbf{L}_{v\iota}^{\mathcal{T}}\mathbf{R}_{\xi\xi}^{-1}\mathbf{L}_{v\iota}\breve{\mathbf{\Sigma}}_{u}^{(k)}(\omega)\right)}{\mathrm{p}} + \epsilon \qquad [2.1.23]$$

***Pseudocode 2: Joint Source Activity and Connectivity***

**INPUTS** $\mathbf{L}_{v\iota}, \mathbf{S}_{vv}(\boldsymbol{\omega}), \mathbf{R}_{\xi\xi}, \{(\mathbb{G}_{\mathcal{i}}, \mathbf{q}_{\mathcal{i}}); \mathcal{i}=\mathbf{1}\cdots\mathbb{q}\}, \mathbf{A}, \mathbf{p}, \mathbf{q}, \mathbf{m}, \boldsymbol{\epsilon}, \boldsymbol{\lambda}(\boldsymbol{\omega})=\sqrt{\boldsymbol{log}(\mathbf{q})/\mathbf{m}}$

**OUTPUTS** $\hat{\mathbf{S}}_{u}(\omega), \widehat{\mathbf{\Theta}}_{u}(\omega), \mathbf{PCoh}(\omega)$

**INITIALIZE** Set: $\hat{\sigma}_{\xi}^{2^{(0)}}=1$, $\widehat{\mathbf{\Theta}}_{u}^{(0)}(\omega)=\mathbf{I}_{\mathrm{q}}$, $\mathbb{K}_{0}^{(0)}=[empty]$

Compute: "SPC" $\breve{\mathbf{\Sigma}}_{u}^{(0)}(\omega)$ [2.1.11], "DSTO" $\breve{\mathbf{T}}_{\iota v}^{(0)}(\omega)$ [2.1.12], "SEC" $\hat{\mathbf{S}}_{u}^{(0)}(\omega)$ [2.1.13], "ESEC" $\breve{\mathbf{\Psi}}_{u}^{(0)}(\omega)$ [2.1.14]

Rescale: "DEC" $\mathbf{S}_{vv}(\omega) \leftarrow \left(\mathrm{q}\,max\left(diag\left(\breve{\mathbf{\Sigma}}_{u}^{(0)}(\omega)\right)\right)/tr\left(\hat{\mathbf{S}}_{u}^{(0)}(\omega)\right)\right)\mathbf{S}_{vv}(\omega)$

**OUTER CYCLE** $k=1\cdots$ (counter)

***step 1*** $\breve{\mathbf{\Sigma}}_{u}^{(k)}(\omega) \leftarrow \left(\widehat{\mathbf{\Theta}}_{u}^{(k-1)}(\omega)=\breve{\mathbf{\Psi}}_{u}^{(k-1)^{-1}}\right)$ Auxiliary quantity "SPC" [2.1.11]

***step 2*** $\left(\breve{\mathbf{\Sigma}}_{u}^{(k)}(\omega)\right)_{\mathbb{K}_{0}^{(k-1)}(\omega)} \leftarrow 0$ Apply sparsity pattern to "SPC"

***step 3*** $\breve{\mathbf{T}}_{\iota v}^{(k)}(\omega)$ Auxiliary quantity "DSTO" [2.1.12]

***step 4*** $\hat{\mathbf{S}}_{u}^{(k)}(\omega)$ Estimator "SEC" [2.1.13]

***step 5*** $\breve{\mathbf{\Psi}}_{u}^{(k)}(\omega)$ Auxiliary quantity "ESEC" [2.1.14]

**INNER CYCLE** $l=1\cdots$ (counter)

***step 6*** $\breve{\mathbf{\Upsilon}}^{(k+1,l)}(\omega)$ Auxiliary quantity "Anatomical Areas Connectivity Selector " [2.1.18]

***step 7*** $\hat{\mathbf{\Gamma}}^{(k+1,l)}(\omega)$ Estimator "Elementwise Connectivity Variances" [2.1.19]

***step 8*** $\widehat{\mathbf{\Theta}}_{u}^{(k+1,l+1)}(\omega)$ Estimator "SP" [2.1.20]

**END**

***step 9*** $\left(\widehat{\boldsymbol{\Theta}}_{\boldsymbol{u}}\right)_{unbiased}^{(k+1)}(\omega)$ Estimator "USP" [2.1.21]

***step 10*** $\mathbb{K}_0^{(k)}(\omega)$ Compute Sparsity Pattern [2.1.22]

***step 11*** $\hat{\sigma}_{\xi}^{2(k+1)}(\omega)$ Estimator "Noise Variance" [2.1.23]

**END**

***step 12*** $\widehat{\mathbf{S}}_{\boldsymbol{u}}(\omega) \leftarrow \widehat{\mathbf{S}}_{\boldsymbol{u}}^{(k)}(\omega)$

***step 13*** $\widehat{\boldsymbol{\Theta}}_{\boldsymbol{u}}(\omega) \leftarrow \left(\widehat{\boldsymbol{\Theta}}_{\boldsymbol{u}}\right)_{unbiased}^{(k)}(\omega)$

***step 12*** $\mathbf{PCoh}(\omega) \leftarrow \widehat{\boldsymbol{\Theta}}_{\boldsymbol{u}}(\omega) \oslash \left(diag\left(\widehat{\boldsymbol{\Theta}}_{\boldsymbol{u}}(\omega)\right) diag\left(\widehat{\boldsymbol{\Theta}}_{\boldsymbol{u}}(\omega)\right)^{\mathcal{T}}\right)$

### 2.2.2 Validation Benchmark

#### *Neuroanatomical and electrophysiological substrate*

The neuroanatomical and electrophysiological substrate for simulation Benchmarking entails elements of a real MEEG scenario. First: T1 MRI image from an individual subject. Second: MEEG Layout information. The T1 MRI image is processed with "FreeSurfer" (https://surfer.nmr.mgh.harvard.edu/), to extract the "Cortex", "Inner Skull", "Outer Skull" and "Scalp" surfaces (head compartments), defined as a high dimensional Mesh (vertices and faces) in the order of tens to hundreds of thousands vertices for the 3 Tesla MRI field, see Figure 3 a) b). A secondary output of FreeSurfer is the Cortical Mesh "Parcellation" into anatomical areas, given a standard atlas in MNI space, i.e. Automated Anatomical Labelling (AAL).

The Brainstorm routines (https://neuroimage.usc.edu/brainstorm/) use FreeSurfer information for the "Head Model" and "Lead Field" construction. It involves associating to the head compartments different conductivities given by experimental information. The Head Model serves as substrate for the solution of "Maxwell Equations" in their stationary approximation "Poisson Equation", of the Scalp Field given to current dipoles at the Cortical Surface. This can be done through different discretization methods, i.e. Finite Element Method (FEM) or Boundary Element Method (BEM), for the projection of a single cortical dipole at the Scalp. In practice the values of the Scalp field are returned at a hundred of Scalp points corresponding to the sensor positions of a specific MEEG Layout (input to Brainstorm). See in Figure 3c the scalp field, at 343 EEG sensors defined into the 10-5 system, due to a single cortical dipole.

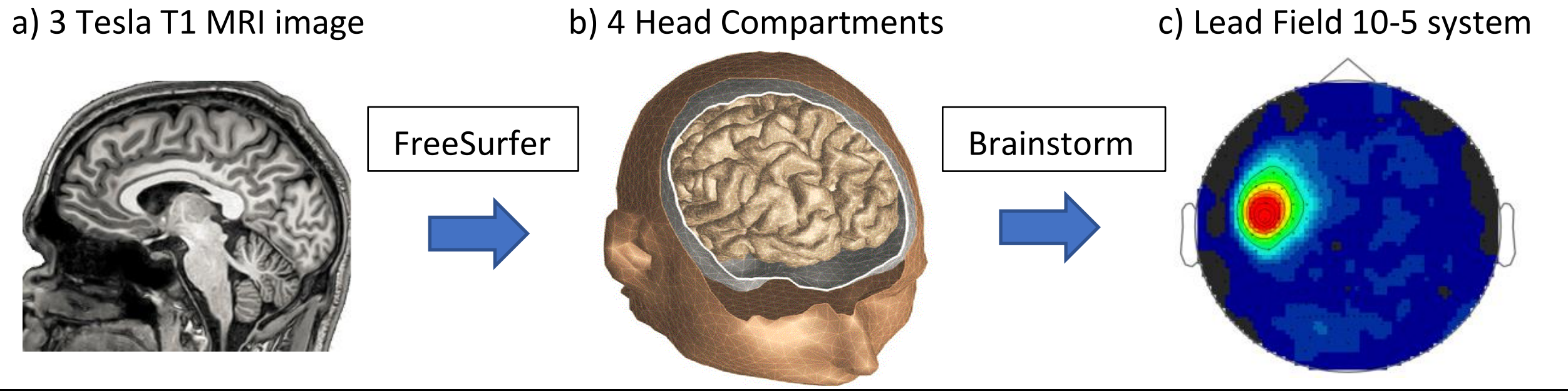

***Figure 3:*** *Simulations substrate derived from Freesurfer and Brainstorm processing for a T1 MRI image and specific MEEG Layout. a) T1 image adquired with a 3T field MRI equipment. b) Surfaces in MNI space, corresponding to 4 Head Compartments (Cortex, Inner Skull, Outer Skull and Scalp). c) Interpolated plot at the 10-5 MEEG sensors space of a SDTO (Lead Field) column, i.e. scalp sensors field of a superior parietal current dipole.*

#### *Sources activity and connectivity set up*

The simulation framework represents a realistic scenario of cortical sources (patches) with variable extensions (geodesic radius) and structured sparse connectivity pattern. The patches centroid location is selected according to distance criterion. **Short-Range**: Defined so that the patches belong to the same anatomical area, without overlap and guaranteeing that the geodesic distance between patches centroid is shorter than 5 cm. **Long-Range**: Defined so that the patches belong to different anatomical area, without overlap and guaranteeing that the geodesic distance between patches centroid is larger than 8 cm. See in Figure 4 an example of three patches.

a) Short-Range

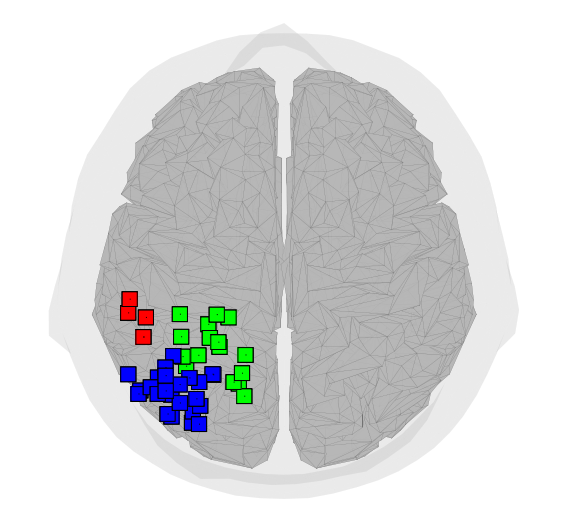

b) Long-Range

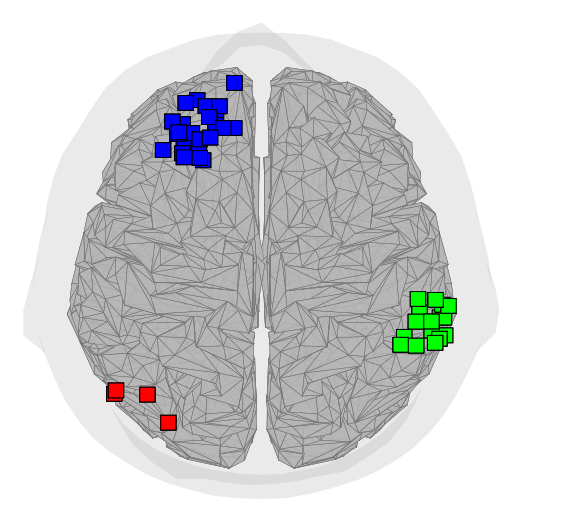

***Figure 4:*** *Example of simulation given by three patches with different extension (4, 8, 24 active generators) for the two citeria of spatial configurations: a) Short-Range instance of Left Occipital (24), Left Superior-Posterior Parietal (8) and Left Inferior-Posterior Parietal (4) active generators patch. b) Long-Range instance of Left Occipital (4), Left Frontal (24) and Right Inferior-Posterior Parietal (8) active generators patch.*

The connectivity structured sparsity patter is given by dense random connections of the sources within the patches and three different modes for the patches interconnections. **Unconnected**: Patches are not connected, i.e. block diagonal structure of the connectivity matrix. **Randomly-Connected**: A random configuration and number of connections take effect, i.e. block structure with several nondiagonal random valued matrix blocks between 1 and (patches#^2 - patches# - 1). **Fully-Connected**: All patches are connected. i.e. full random valued matrix. See in Figure 5 an example of the three connectivity modes defined on three patches.

a) Unconnected

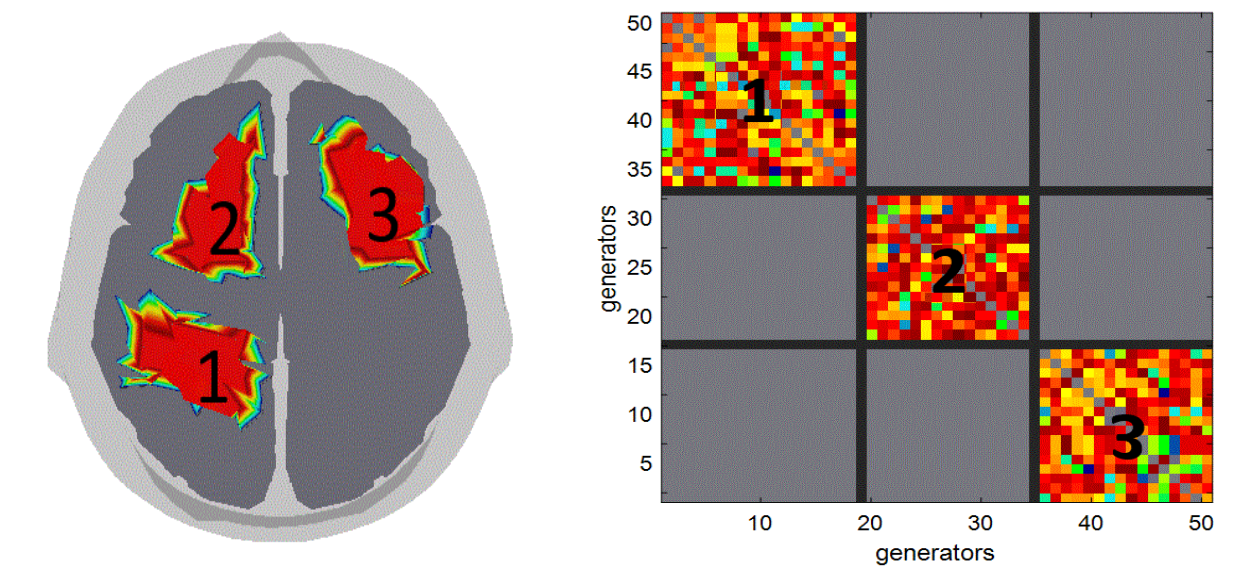

b) Randomly-Connected (1<->3)

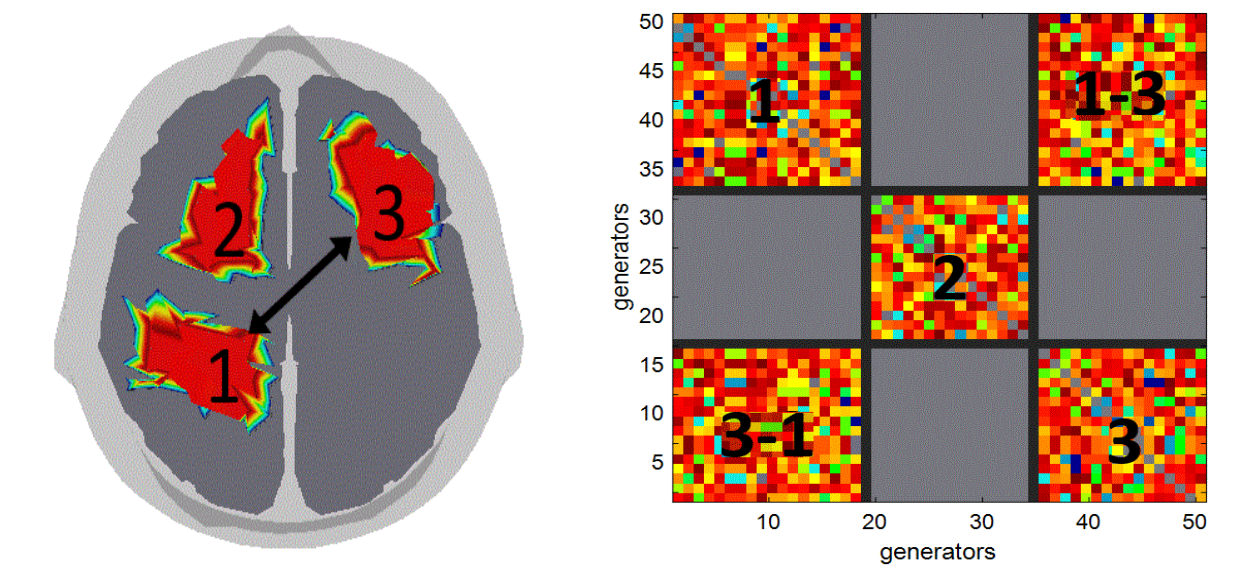

c) Fully-Connected

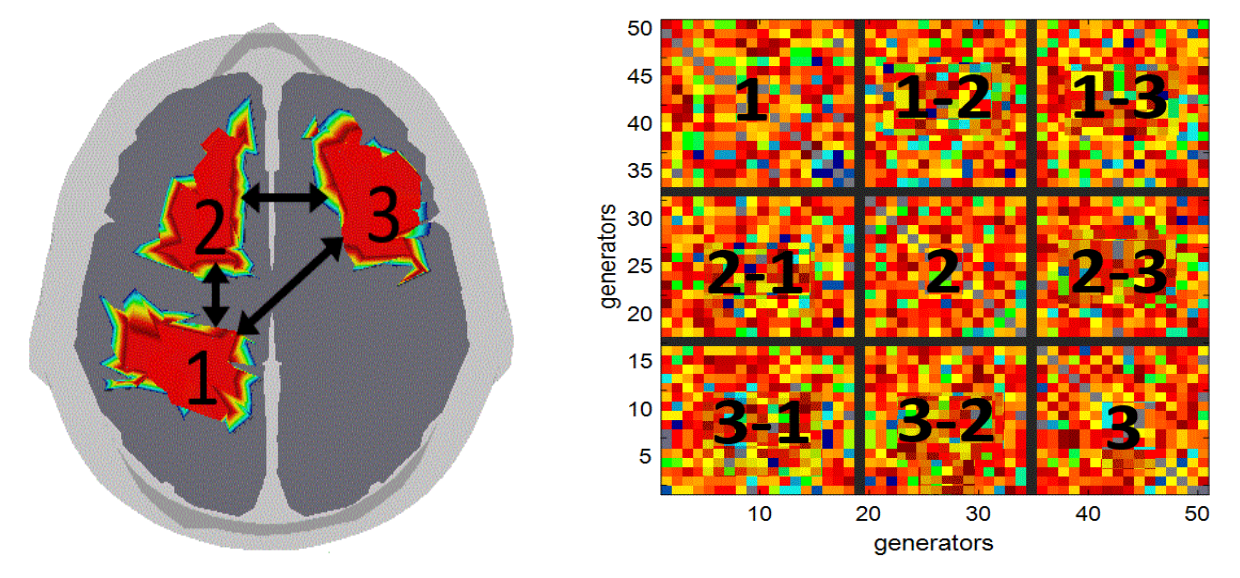

d) Randomly-Connected (1<->2) (1<->3)

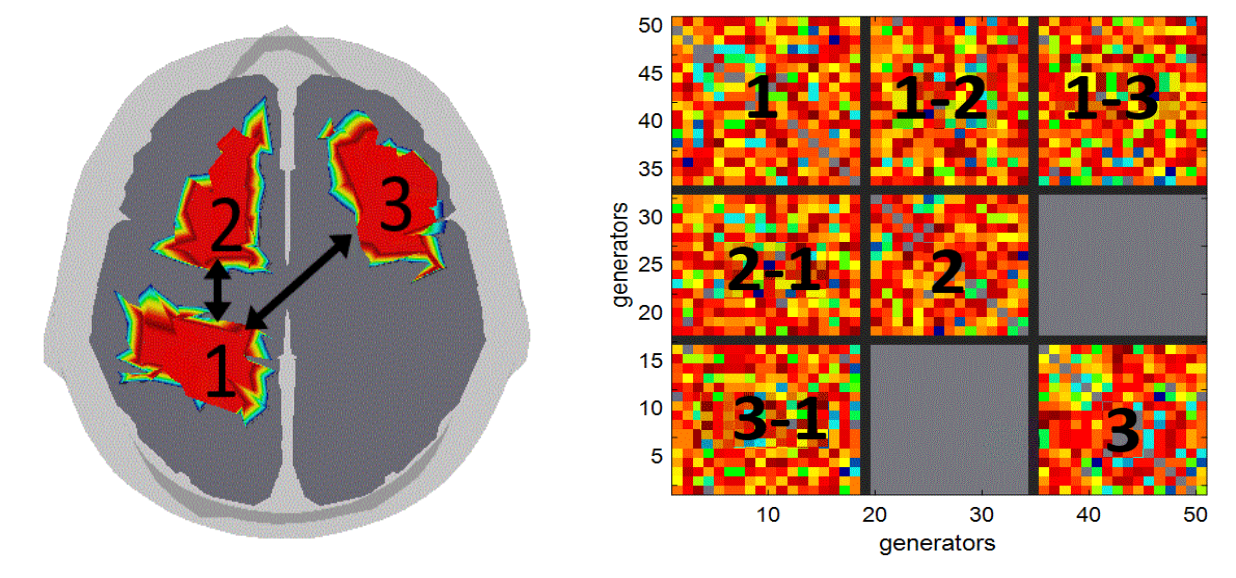

***Figure 5:*** *Three connectivity modes ilustrated with an schematic of different active patches "1,2,3" at the cortical surface and 2D plots of the precision matrix elements corresponding to the active patches. a) Unconnected. c) Randomly-Connected "1<->3". b) Fully-Connected. d) Randomly-Connected "1<->2" "1<->3"*

***Simulation benchmarking implementation***

The simulations are based on a synthetic source precision matrix $\mathbf{\Theta}_{\boldsymbol{\iota}}^{(sim)}$, which is used to generate samples of synthetic source activity samples $\left(\boldsymbol{\iota}_m^{(sim)}, m = 1 \ldots \mathrm{m}\right)$, subsequently projected to the scalp by the SDTO $\mathbf{L}_{\boldsymbol{v\iota}}^{(sim)}$ and mixed with noise samples $\left(\boldsymbol{\xi}_m^{(sim)}, m = 1 \ldots \mathrm{m}\right)$. These elements conform synthetic Data samples $\left(\boldsymbol{v}_m^{(sim)}, m = 1 \ldots \mathrm{m}\right)$ and its corresponding DEC matrix $\mathbf{S}_{\boldsymbol{vv}}^{(sim)}$. See Pseudocode 3.

To avoid the Inverse Crime the STDO $\mathbf{L}_{\boldsymbol{v\iota}}^{(sim)}$ used for simulation purpose is different than the SDTO $\mathbf{L}_{\boldsymbol{v\iota}}$ used for reconstruction. It thus involves the processing by the scheme of Figure 3 of an additional T1 image for the simulations, hereinafter denoted $\mathrm{T1}^{(sim)}$.

The source precision matrix, in the subspace of active generators $\mathbb{G}_{(sim)}$ with size denoted $\mathrm{q}_{(sim)}$, is a hermitic positive definite block array of complex numbers. Creating this array is done in four steps. First: Generating its Real symmetric and Imaginary antisymmetric parts, i.e. $\Re\left(\mathbf{\Theta}_{\boldsymbol{\iota}}^{(sim)}\right)$ and $\Im\left(\mathbf{\Theta}_{\boldsymbol{\iota}}^{(sim)}\right)$, by gaussian random entries. Second: Compositing $\mathbf{\Theta}_{\boldsymbol{\iota}}^{(sim)}$ by its Real and Imaginary parts. Third: Equating to zero those elements where no connections were defined, i.e. set of matrix indices $\mathbb{K}_0^{(sim)}$ in the subspace $\mathbb{G}_{(sim)}$. Fourth: Correcting the positive definiteness of $\mathbf{\Theta}_{\boldsymbol{\iota}}^{(sim)}$.

The noise samples are given by two components. First: Sensors (instrumentation) noise samples $\left(\boldsymbol{\xi}_m^{(inst)}, m = 1 \ldots \mathrm{m}\right)$, given by complex vectors composited of real an imaginary gaussian random entries. Second: Sources (biological) noise samples $\left(\boldsymbol{\xi}_m^{(biol)}, m = 1 \ldots \mathrm{m}\right)$, given by complex vectors composited of real an imaginary gaussian random entries, to be projected at the scalp sensors by the SDTO $\mathbf{L}_{\boldsymbol{v\iota}}^{(sim)}$. Both noise components are normalized by its energy and scaled to a fraction of the synthetic Data energy, by means of a user defined noise ratio coefficient $\sigma_{\xi}^{(sim)}$.

***Pseudocode 3: Simulations***

**INPUTS** $\mathbf{L}_{\boldsymbol{v\iota}}^{(sim)}$, $\mathbb{G}_{(sim)}$, $\mathbb{K}_0^{(sim)}$, $\mathbf{p}$, $\mathbf{q}_{(sim)}$, $\mathbf{m}$, $\boldsymbol{\sigma}_{\xi}^{(sim)}$

**OUTPUTS** $\mathbf{S}_{\boldsymbol{vv}}^{(sim)}$, $\mathbf{S}_{\boldsymbol{\iota}}^{(sim)}$, $\mathbf{\Theta}_{\boldsymbol{\iota}}^{(sim)}$

**CONNECTIVITY**

***step 1*** $\Re\left(\mathbf{\Theta}_{\boldsymbol{\iota}}^{(sim)}\right) \leftarrow randn\left(\mathbf{q}_{(sim)}, \mathbf{q}_{(sim)}\right)$ Real part in subspace $\mathbb{G}_{(sim)}$

***step 2*** $\Re\left(\mathbf{\Theta}_{\boldsymbol{\iota}}^{(sim)}\right) \leftarrow \left(\Re\left(\mathbf{\Theta}_{\boldsymbol{\iota}}^{(sim)}\right) + \Re\left(\mathbf{\Theta}_{\boldsymbol{\iota}}^{(sim)}\right)^{\mathcal{T}}\right)/2$ Symmetric property

***step 3*** $\Im\left(\mathbf{\Theta}_{\boldsymbol{\iota}}^{(sim)}\right) \leftarrow randn\left(\mathbf{q}_{(sim)}, \mathbf{q}_{(sim)}\right)$ Imaginary part in subspace $\mathbb{G}_{(sim)}$

***step 4*** $\Im\left(\mathbf{\Theta}_{\boldsymbol{\iota}}^{(sim)}\right) \leftarrow \left(\Im\left(\mathbf{\Theta}_{\boldsymbol{\iota}}^{(sim)}\right) - \Im\left(\mathbf{\Theta}_{\boldsymbol{\iota}}^{(sim)}\right)^{\mathcal{T}}\right)/2$ Antisymmetric property

***step 5*** $\mathbf{\Theta}_{\boldsymbol{\iota}}^{(sim)} \leftarrow complex\left(\Re\left(\mathbf{\Theta}_{\boldsymbol{\iota}}^{(sim)}\right), \Im\left(\mathbf{\Theta}_{\boldsymbol{\iota}}^{(sim)}\right)\right)$ Complex compositing

***step 6*** $\mathbf{\Theta}_{\boldsymbol{\iota}}^{(sim)}\left(\mathbb{K}_0^{(sim)}\right) \leftarrow 0$ Set to zero elements in unconnected indexes

***step 7*** *If* $min\left(svds\left(\mathbf{\Theta}_{\boldsymbol{\iota}}^{(sim)}\right)\right) \leq 0$ then correct singular values to be greater than zero

**ACTIVITY**

***step 8*** $\left(\boldsymbol{\iota}_m^{(sim)}\right)_{\mathbb{G}_{(sim)}} \sim N_{\mathbf{q}_{(sim)}}^{\mathbb{C}}\left(\left(\boldsymbol{\iota}_m^{(sim)}\right)_{\mathbb{G}_{(sim)}} \middle| \mathbf{0}, \mathbf{\Theta}_{\boldsymbol{\iota}}^{(sim)^{-1}}\right); m = 1 \ldots \mathrm{m}$

***step 9*** $\mathbf{S}_{\boldsymbol{\iota}}^{(sim)} \leftarrow \frac{1}{\mathrm{m}} \sum_{m=1}^{\mathrm{m}} \boldsymbol{\iota}_m^{(sim)} \boldsymbol{\iota}_m^{(sim)\dagger}$ Synthetic "SEC"

**NOISE**

***step 10*** $\boldsymbol{\xi}_m^{(inst)} \sim randn(\mathrm{p}, 1); m = 1 \ldots \mathrm{m}$ Instrumental Noise

***step 11*** $\mathbf{S}_{\boldsymbol{\xi\xi}}^{(inst)} \leftarrow \frac{1}{\mathrm{m}} \sum_{m=1}^{\mathrm{m}} \boldsymbol{\xi}_m^{(inst)} \boldsymbol{\xi}_m^{(inst)\dagger}$ Instrumental Noise Empirical Covariance **step 12** $\boldsymbol{\xi}_m^{(biol)} \sim randn(\mathrm{q},1);\ m = 1 \dots \mathrm{m}$ Biological Noise at Generator Space

**step 13** $\boldsymbol{\xi}_m^{(biol)} \leftarrow \mathbf{L}_{\boldsymbol{v\iota}}^{(sim)} \boldsymbol{\xi}_m^{(biol)};\ m = 1 \dots \mathrm{m}$ Biological Noise at Sensor Space

**step 14** $\mathbf{S}_{\boldsymbol{\xi\xi}}^{(biol)} \leftarrow \frac{1}{\mathrm{m}} \sum_{m=1}^{\mathrm{m}} \boldsymbol{\xi}_m^{(biol)} \boldsymbol{\xi}_m^{(biol)\dagger}$ Biological Noise Empirical Covariance

**SCALING**

**step 15** $\mathbf{S}_{\boldsymbol{vv}}^{(sim)} \leftarrow \mathbf{L}_{\boldsymbol{v\iota}}^{(sim)} \mathbf{S}_{\boldsymbol{\iota}}^{(sim)} \mathbf{L}_{\boldsymbol{v\iota}}^{(sim)\mathcal{T}}$ Synthetic Ideal "DEC"

**step 16** $\boldsymbol{\xi}_m^{(inst)} \leftarrow \boldsymbol{\xi}_m^{(inst)} tr\left(\mathbf{S}_{\boldsymbol{vv}}^{(sim)}\right) / \left(\sigma_{\xi}^{(sim)} tr\left(\mathbf{S}_{\boldsymbol{\xi\xi}}^{(inst)}\right)\right);\ m = 1 \dots \mathrm{m}$ Instrumental Noise

**step 17** $\boldsymbol{\xi}_m^{(biol)} \leftarrow \boldsymbol{\xi}_m^{(biol)} tr\left(\mathbf{S}_{\boldsymbol{vv}}^{(sim)}\right) / \left(\sigma_{\xi}^{(sim)} tr\left(\mathbf{S}_{\boldsymbol{\xi\xi}}^{(biol)}\right)\right);\ m = 1 \dots \mathrm{m}$ Biological Noise

**step 18** $\boldsymbol{\xi}_m^{(sim)} \leftarrow \boldsymbol{\xi}_m^{(inst)} + \boldsymbol{\xi}_m^{(biol)};\ m = 1 \dots \mathrm{m}$ Total Noise

**DATA**

**step 19** $\boldsymbol{v}_m^{(sim)} = \mathbf{L}_{\boldsymbol{v\iota}}^{(sim)} \boldsymbol{\iota}_m^{(sim)} + \boldsymbol{\xi}_m^{(sim)},\ m = 1 \dots \mathrm{m}$ Synthetic Noisy Data

**step 20** $\mathbf{S}_{\boldsymbol{vv}}^{(sim)} = \frac{1}{\mathrm{m}} \sum_{m=1}^{\mathrm{m}} \boldsymbol{v}_m^{(sim)} \boldsymbol{v}_m^{(sim)\dagger}$ Synthetic Noisy DEC

**END**

***Quality Measures (EMD, CEMD)***

A measure representative of distortion in general scenarios is the Earth Movers' Distance, for both the typical vector case (EMD) and its cartesian extension to hermitic matrices (CEMD). In the State of the Art of Inverse Solution the EMD has been stablished as the most sensitive when compared to other quality measures, i.e. typical Dipole Localization Error or Binary Classification, i.e. Receiving Operating Characteristic, Precision, Recall and F1. The EMD is applied here to measure the distortion of the source activity estimates, represented by the diagonal elements of the SEC $diag\left(\hat{\mathbf{S}}_{\boldsymbol{\iota}}(\omega)\right)$, from the diagonal of the simulated SEC $diag\left(\mathbf{S}_{\boldsymbol{\iota}}^{(sim)}\right)$, i.e. EMD-SEC. Meanwhile, the CEMD measures the distortion of the estimated connectivity $\widehat{\boldsymbol{\Theta}}_{\boldsymbol{\iota}}(\omega)$ from the simulated one $\boldsymbol{\Theta}_{\boldsymbol{\iota}}^{(sim)}$, i.e. CEMD-SP.

### 2.2.3 Toolbox Technical Route

BC-VARETA technical route possess two main branches, see diagram in Figure 6. First (left branch): MEEG Data processing, through the Third Generation method BC-VARETA, given the frequency domain representation of and individual subject time series (inputs: "MEEG Layout", "MEEG Data" and "T1" MRI, Prior Connectivity Maps "PCM"; outputs: "SEC", "USP" and "PCOH" ). Second (right branch): Validation Benchmark for Third Generation methods, merged here to BC-VARETA routines (inputs: "MEEG Layout" and two T1 MRI for estimation "T1$^{(est)}$" and simulation "T1$^{(sim)}$"; outputs: "SEC", "USP", "PCOH", "EMD-SEC", CEMD-USP and "CEMD-PCOH"). The first release of our toolbox (BC-VARETA 1.0) is freely availlable in matlab format at the GitHub link: https://github.com/egmoreira/BC-VARETA-toolbox

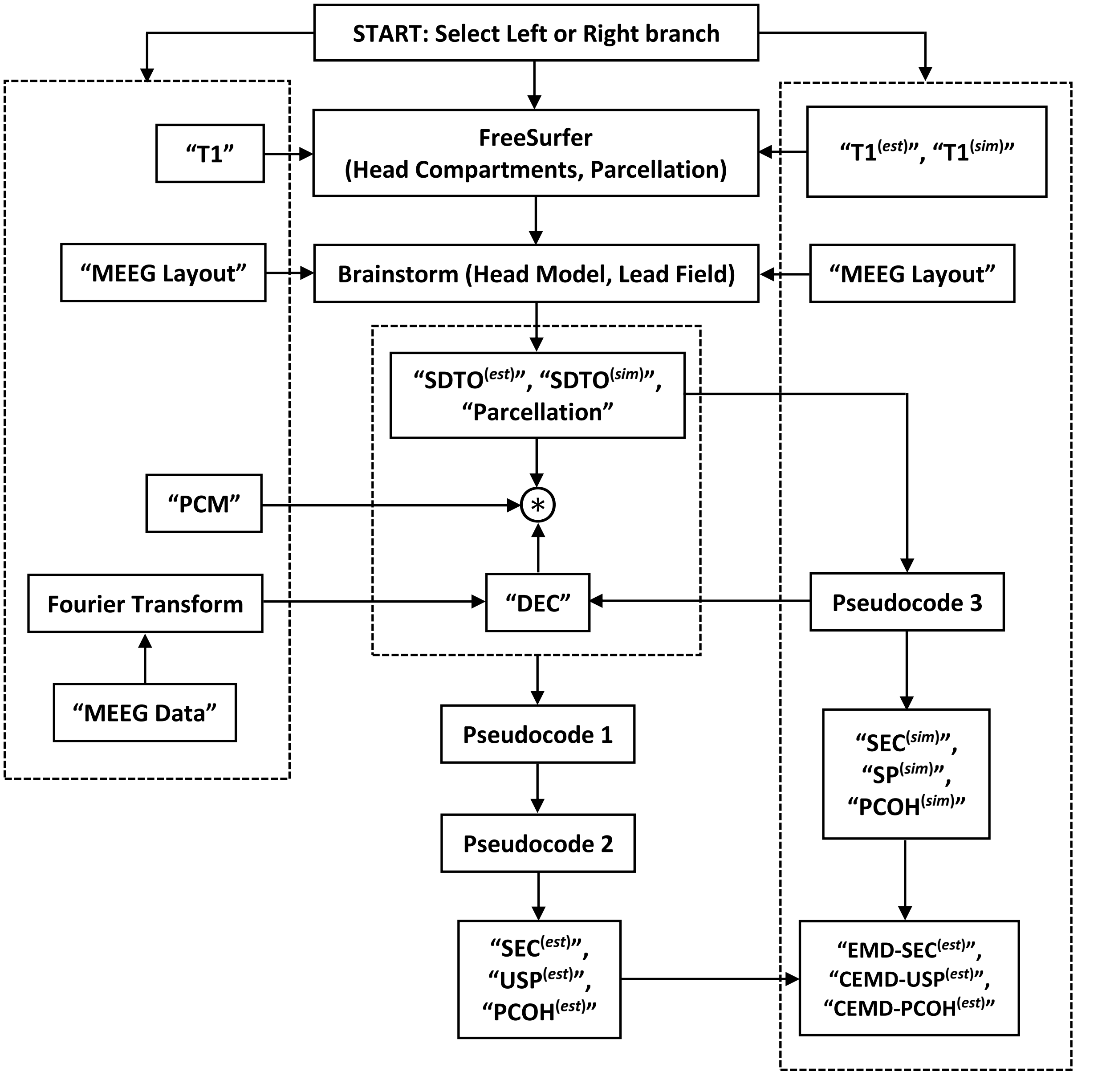

***Figure 6:*** *Block diagram (technical route) of BC-VARETA 1.0 for both MEEG Data processing (left branch) and Validation Benchmark (right branch).*

# 3 RESULTS AND DISCUSSION

## 3.1 Simulation Analysis

For an evaluation of BC-VARETA inference framework, at each condition specified by the "Validation Benchmark", i.e. distance (short-range and long range) and connectivity modes (unconnected, randomly-connected and fully-connected), we generated 100 random trials of patches centroids. The sources were set up similarly to the example in Figure 4, so that every random trial included three patches with spatially distributed activity over (4,8,24) nodes correspondingly. Also, the connectivity modes were set in correspondence to what was illustrated in Figure 5. The synthetic DEC was obtained from 400 samples of simulated source activity and noise, following the procedure described in Pseudocode 3. The signal to noise ratio was adjusted to four different levels none (∞dB), low (19dB), middle (7dB) and high (0dB).

For a complementary quantitative analysis, we considered some other quality measures of the difference between the simulated SEC diagonal its reconstruction. We implemented five measures derived from Receiver Operational Characteristic (ROC), i.e. Sensitivity (TPR), Specificity (TNR), Area Under ROC Curve (AUC) and F1 score (F1S). Also, a homogeneity test was applied to the simulated and reconstructed SP matrix, i.e. M Box statistical test (*Pituch and Stevens, 1994*). It serves to evaluate in general scenarios the similarity between precision or covariance matrices ($\boldsymbol{\Sigma}_0, \boldsymbol{\Sigma}_1$) by returning the rejection rate of the hypothesis ($\mathrm{H}_0 : \boldsymbol{\Sigma}_0 = \boldsymbol{\Sigma}_1$).

For comparison purpose we use well stablished Second Generation MEEG methods: Exact Low-Resolution Tomography (eLORETA) and Linearly Constrained Minimum Variance (LCVM). eLORETA (*Pascual, 2002*) constitutes the most stablished and robust (provides zero localization error in case of a single source reconstruction). The SEC is extracted from the sources estimates after convergence, like in formula [2.1.13], and the connectivity from the as the inverse of the source covariance given by a statistical formula. LCMV (*Van Veen et al. 1997*), also well stablished but among the family of Beam Former methods, constitutes a qualitative different approach when compared to eLORETA. That enriches our validation with a higher contrast of the results. Roughly, it consists on the Spatial Filtering of the Forward Equation [2.1.1]. The LCMV focuses only in the Sources' Variances, ruling out the Covariance structure from the First Level of Inference, but it also provides a Connectivity analysis through statistical formulas of the sources covariance.

### 3.1.1 Source Activity Reconstruction

In Figure 7 we show, for eLORETA, LCMV and BC-VARETA, binary plots of the reconstructed source activity on the cortical surface, obtained by two simulated instances in unconnected mode and at ∞dB noise level. The reconstructed activity with eLORETA spreads over the whole cortical surface, reflecting the largest number of false positives, when compared to LCMV and BC-VARETA, and making difficult to distinguish the three main activations (patches). LCMV reconstruction is in change more accurate since most of the positives centered around the true activations, but still too smooth and revealing spatially distributed activity that oversteps the simulated one.

BC-VARETA reconstruction was the most sparse and accurate. Being qualitative different to the previous solutions, it identifies with high accuracy the presence of three main activations alone, extending only over the cortical areas where the activity was simulated. This situation holds in either long-range and short-range condition, demonstrating qualitatively the ability of our method retrieving variable degrees of sparsity in spatially distributed activity. This, being claimed in the pass by VARETA method, is reinforced now with our model assumptions and sophisticated priors.

The localization performance was quantified by ROC and EMD measures, reported as the average and standard deviation for 100 trials of the two instances, see Table 1. According with all quality measures the highest performance was achieved by BC-VARETA method, consistently to what was discussed in the qualitative analysis above. eLORETA exhibited the poorest performance, also in correspondence to the qualitative analysis, due to large number of false positives of its reconstructed sources. LCMV reconstruction achieved an intermediate number of false positives, between eLORETA and BC-VARETA results, reflected by the ROC analysis and consistently to the EMD values.

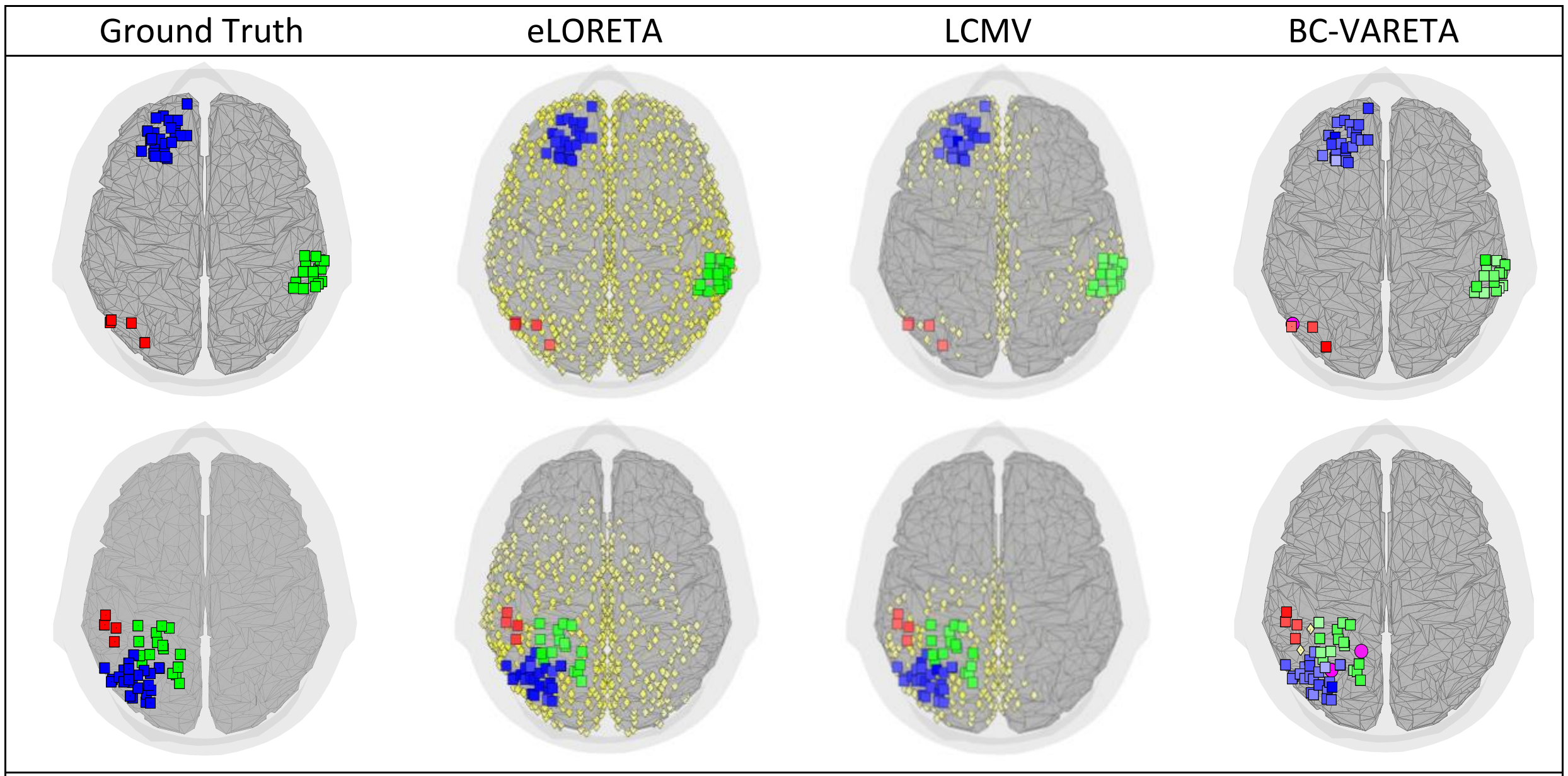

***Figure 7:*** *Binary cortical maps of three patches with different size, i.e. 4 (red), 8 (green), 24 (blue) active generators, in two instances of long range (top row) and short-range (bottom-row) configurations. From left to right Ground Truth, reconstruction with eLORETA, LCMV and BC-VARETA. The squares represent the True Positive estimation, yellow diamonds represent the False Positive estimation and the magenta circle represents the False Negative estimation. This instances where generated in unconnected mode and the noise level was set to ∞dB.*

***Table 1:*** *ROC and EMD measures in source localization performance for the methods: eLORETA, LCMV and BC-VARETA, computed in 100 trials of both instances: short-range (top row) and long-range (bottom row). The instances were generated analogously to those of Figure 7.*

| **Long-range instances** | | | | | |
|---|---|---|---|---|---|
| **Method** | TPR | TNR | AUC | F1 | SEC-EMD |
| **eLORETA** | 1.0 ± 0.00 | 0.10 ± 0.05 | 0.12 ± 0.05 | 0.03 ± 0.00 | 19.70 ± 1.96 |
| **LCMV** | 1.0 ± 0.00 | 0.53 ± 0.09 | 0.54 ± 0.09 | 0.07 ± 0.01 | 19.51 ± 4.15 |
| **BC-VARETA** | 0.95 ± 0.06 | 1.00 ± 0.00 | 0.99 ± 0.00 | 0.97 ± 0.04 | 2.65 ± 0.61 |
| **Short-range instances** | | | | | |
| **Method** | TPR | TNR | AUC | F1 | SEC-EMD |
| **eLORETA** | 1.00 ± 0.00 | 0.31 ± 0.07 | 0.33 ± 0.07 | 0.05 ± 0.00 | 19.17 ± 4.11 |
| **LCMV** | 1.00 ± 0.00 | 0.67 ± 0.05 | 0.68 ± 0.05 | 0.10 ± 0.01 | 18.95 ± 3.41 |
| **BC-VARETA** | 0.95 ± 0.05 | 0.99 ± 0.00 | 0.99 ± 0.00 | 0.97 ± 0.03 | 3.08 ± 0.662 |

### 3.1.2 Connectivity Reconstruction

The connectivity retrieval performance was evaluated for the short-range configuration of Figure 7, for 4 connectivity instances: unconnected, randomly-connected 1<->3, randomly-connected 1<->2 1<->3 and fully-connected. See in Figure 8 bidimensional plots of eLORETA, LCMV and BC-VARETA reconstructed connectivity, shown as block matrices in the reduced space of simulated sources.

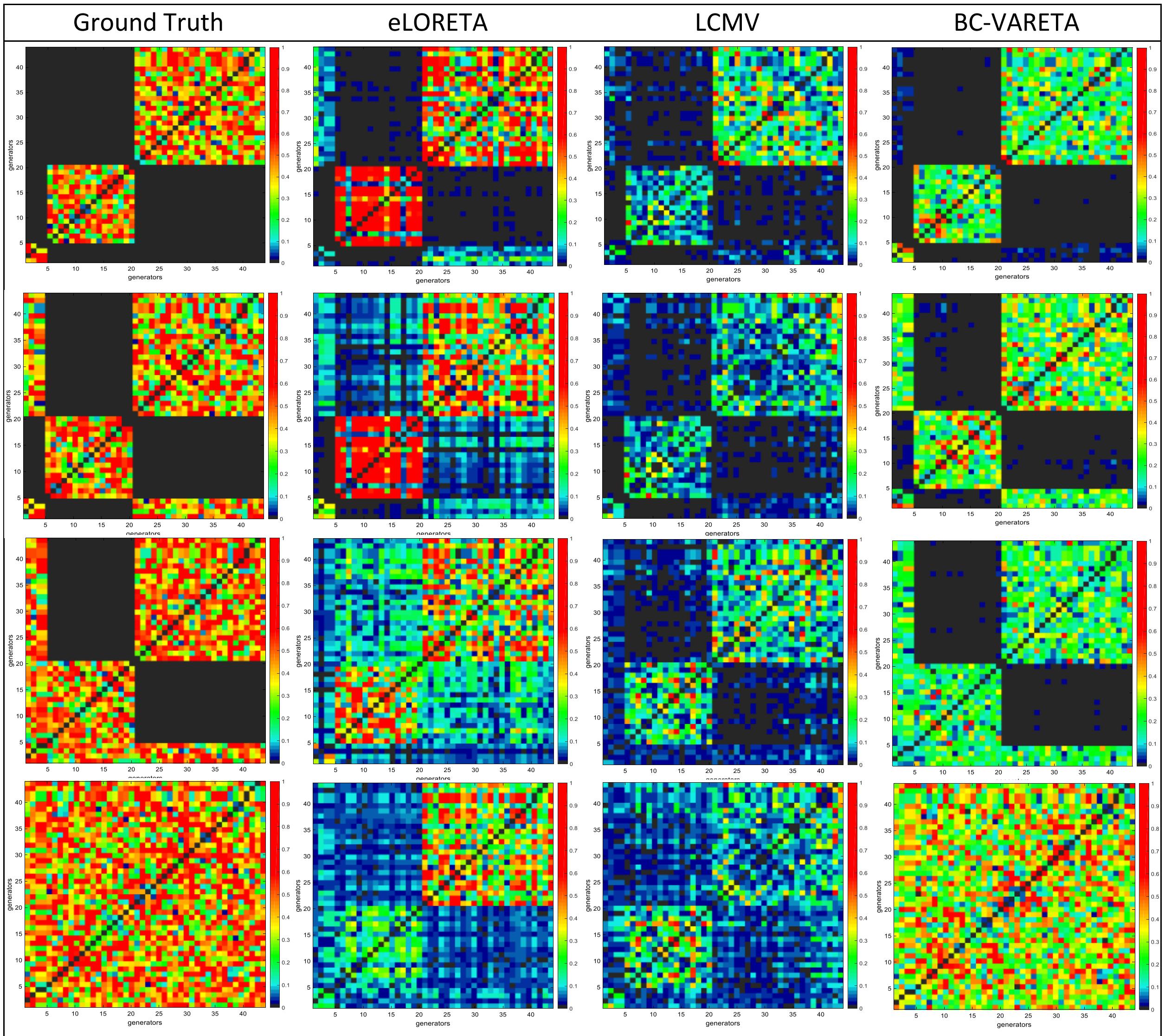

***Figure 8:*** *Bidimensional plots (matrix) of the three patches connectivity, in the short-range configuration corresponding to Figure 6, for 4 connectivity instances: unconnected (first row), randomly-connected 1<->3 (second row), randomly-connected 1<->2 1<->3 (third row) and fully connected (fourth row). From left to right cortical maps of Ground Truth, reconstruction with eLORETA, LCMV and BC-VARETA.*

The reconstructed connectivity with eLORETA spreads over the unconnected blocks, revealing that the true positives associated to source localization are given by the overestimation (false positives) of adjacent patches. This makes extremely difficult to judge the actual performance of this method, even in what appears to be the most straightforward instance: unconnected mode. In the connectivity scenario LCMV exhibit a performance which qualitatively does not distinguish from eLORETA reconstruction. Even when the number of connectivity false positives is slightly diminished, still reflects overlap of the patches estimates.

Again, at the connectivity level, BC-VARETA reconstruction was the most accurate, showing in comparison to eLORETA and LCMV just a few false positives of that are negligible to judge by the color scale of these bidimensional plots. This situation holds in either of the connectivity modes, that are used in Figure 8 to illustrate the connectivity retrieval performance. It confirms our expectations about BC-VARETA model assumptions, i.e. the effect of the sparse connectivity model, in a situation where the structured sparsity pattern built on variable extent of the matrix blocks representing the patches interconnections.

The connectivity reconstruction performance was quantified by M Box statistical test and CEMD measures in 100 trials of the short-range configuration, see Table 2, for each of the instances (connectivity modes) in Figure 8. According to all quality measures the highest performance in connectivity reconstruction was achieved by BC-VARETA method, this was consistent to the qualitative analysis at the connectivity level and the source reconstruction study of the previous section. eLORETA and LCMV performance was again poorer and characterized by a narrow difference of the average between both methods measures, in comparison to the standard deviation, also in correspondence to what was discussed above about their qualitatively similar connectivity maps.

***Table 2:*** *M Box statistical test and Cartesian Earth Movers' Distance measures in connectivity reconstruction performance for the methods: eLORETA, LCMV and BC-VARETA. Computed, analogously to Figure 8, for 100 trials of the short-range configuration and three connectivity modes: unconnected, randomly-connected and fully-connected.*

| **Connectivity reconstruction performance based on Box test** | | | | |
|---|---|---|---|---|
| | **Connectivity type** | | | |
| **Method** | unconnected | randomly-connected (1<->3) | randomly-connected (1<->2) (1<->3) | fully-connected |
| **eLORETA** | 6.450 ± 0.295 | 7.244 ± 0.306 | 7.598 ± 0.168 | 9.954 ± 0.034 |
| **LCMV** | 6.786 ± 0.193 | 7.683 ± 0.229 | 7.810 ± 0.182 | 9.849 ± 0.058 |
| **BC-VARETA** | 6.537 ± 0.577 | 6.475 ± 0.384 | 6.503 ± 0.137 | 7.291 ± 0.471 |
| **Connectivity reconstruction performance based on Cartesian Earth Movers' Distance (CEMD-USP)** | | | | |
| | **Connectivity type** | | | |
| **Method** | unconnected | randomly-connected (1<->3) | randomly-connected (1<->2) (1<->3) | fully-connected |
| **eLORETA** | 277.04 ± 32.16 | 269.49 ± 31.79 | 267.79 ± 33.42 | 273.8 ± 37.78 |
| **LCMV** | 309.35 ± 22.90 | 321.70 ± 15.28 | 302.09 ± 21.00 | 303.6 ± 23.76 |
| **BC-VARETA** | 93.02 ± 7.93 | 117.80 ± 8.54 | 131.39 ± 12.74 | 107.7 ± 13.91 |

### 3.1.3 Effect of Noise Level in Reconstruction Performance

The effect of noise was studied by replicating the previous simulations with different values of SNR level, i.e. low (19 dB), middle (7dB) and high (0dB). Figure 9 shows an example of three patches in the long-range and unconnected condition. From this qualitative analysis it can be noticed that the source activity and connectivity reconstruction with the all methods deteriorates as the SNR level decreases. However, BC-VARETA outperform eLORETA and LCMV in all scenarios, consistently to the results at $\infty$dB of the previous section. Furthermore, we studied the source localization performance in 100 trial of the long-range instance and unconnected by ROC and EMD measures, see Table 3. Also, we explored their behavior for the different connectivity modes in the long-range configuration by M Box and CEMD measures, see Table 4. This quantitative analysis confirmed the point discussed above, on the deterioration of the results with SNR decrease, being BC-VARETA performance the robust for all noise levels.

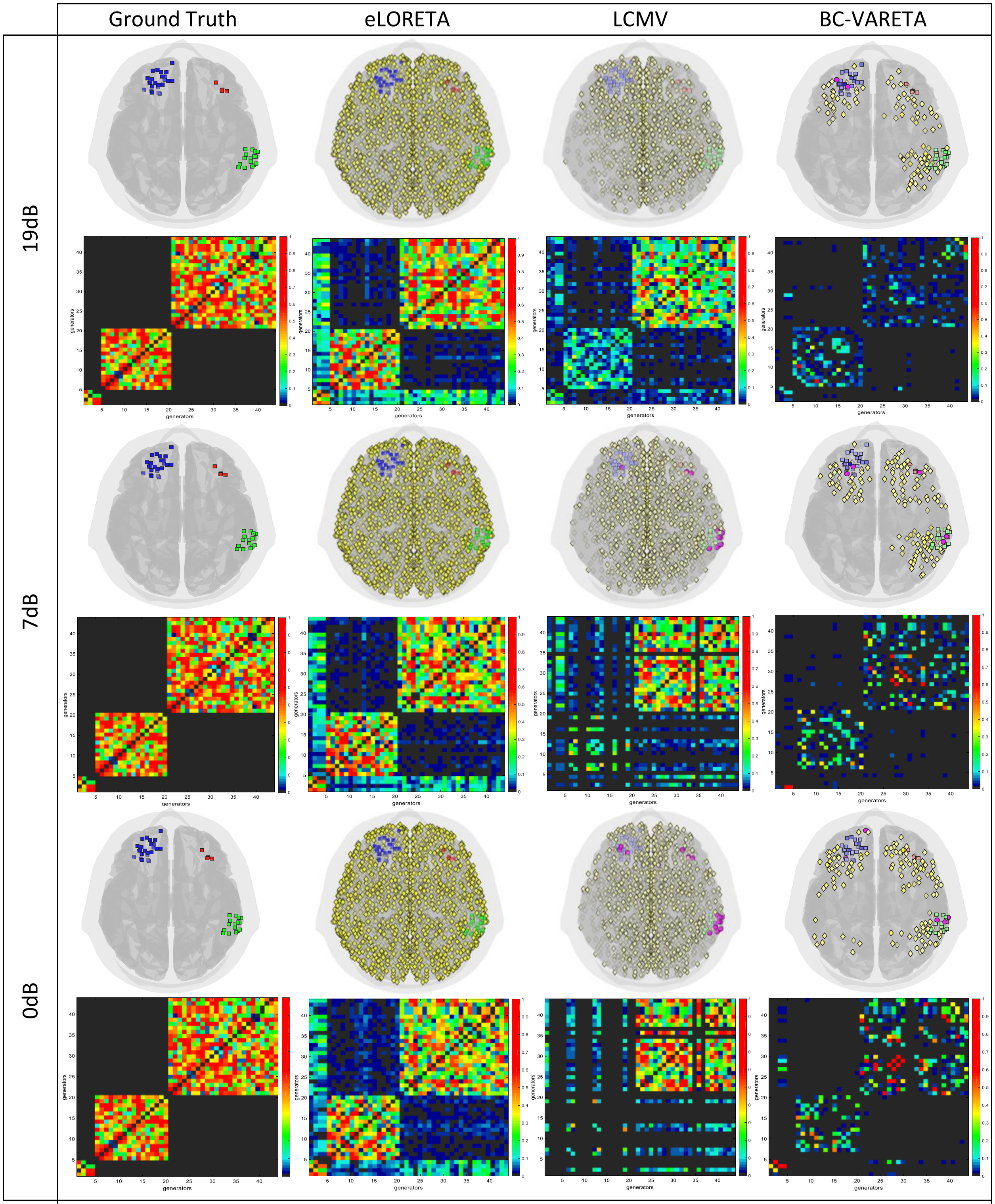

***Figure 9**. Binary cortical maps of three patches with different size, i.e. 4 (red), 8 (green), 24 (blue) active generators, in long range configurations. From left to right Ground Truth, reconstruction with eLORETA, LCMV and BC-VARETA. The squares represent the True Positive estimation, yellow diamonds represent the False Positive estimation and the magenta circle represents the False Negative estimation. This instances where generated in unconnected mode for different noise levels: 19dB, 7dB and 0dB.*

***Table 3:*** *ROC and EMD measures in source localization performance under different levels of noise (19dB, 7dB and 0dB) for the methods: eLORETA, LCMV and BC-VARETA, computed in 100 trials of the long-range configuration and unconnected mode.*

| **Source reconstruction performance based on ROC analysis and Earth Mover's Distance** | | | | | | |
|---|---|---|---|---|---|---|
| **SNR** | Method | TPR | TNR | AUC | F1 | EMD-SEC |
| **19 dB** | eLORETA | 1.00 ± 0.00 | 0.00 ± 0.00 | 0.02 ± 0.00 | 0.02 ± 0.00 | 44.3 ± 3.02 |
| | LCMV | 0.98 ± 0.01 | 0.36 ± 0.04 | 0.37 ± 0.04 | 0.06 ± 0.00 | 25.4 ± 3.43 |
| | BC-VARETA | 0.84 ± 0.06 | 0.93 ± 0.01 | 0.92 ± 0.01 | 0.36 ± 0.06 | 11.4 ± 4.58 |
| **7 dB** | eLORETA | 1.00 ± 0.00 | 0.00 ± 0.00 | 0.02 ± 0.00 | 0.04 ± 0.00 | 43.1 ± 1.91 |
| | LCMV | 0.85 ± 0.07 | 0.32 ± 0.04 | 0.33 ± 0.04 | 0.05 ± 0.04 | 26.8 ± 1.77 |
| | BC-VARETA | 0.74 ± 0.08 | 0.92 ± 0.01 | 0.91 ± 0.01 | 0.28 ± 0.04 | 14.2 ± 3.85 |
| **0 dB** | eLORETA | 1.00 ± 0.00 | 0.00 ± 0.00 | 0.01 ± 0.00 | 0.03 ± 0.00 | 41.7 ± 1.91 |
| | LCMV | 0.72 ± 0.12 | 0.29 ± 0.04 | 0.30 ± 0.04 | 0.03 ± 0.00 | 25.2 ± 2.54 |
| | BC-VARETA | 0.65 ± 0.11 | 0.92 ± 0.01 | 0.91 ± 0.01 | 0.22 ± 0.04 | 12.9 ± 3.18 |

***Table 4:*** *M Box statistical test and Cartesian Earth Movers' Distance measures in connectivity reconstruction performance under different levels of noise (19dB, 7dB and 0dB) for the methods: eLORETA, LCMV and BC-VARETA. Computed, analogously to Figure 9, in 100 trials of the long-range configuration and three connectivity modes: unconnected, randomly-connected and fully-connected.*

| **Connectivity performance based on Box's M test statistic** | | | | | |
|---|---|---|---|---|---|
| | | **Connectivity type** | | | |
| **SNR** | Method | unconnected | randomly-connected (1<->3) | randomly-connected (1<->2) (1<->3) | fully-connected |
| **19 dB** | eLORETA | 6.471 ± 0.304 | 7.216 ± 0.287 | 7.621 ± 0.224 | 9.950 ± 0.037 |
| | LCMV | 6.506 ± 0.173 | 6.820 ± 0.178 | 7.660 ± 0.146 | 9.883 ± 0.046 |
| | BC-VARETA | 6.105 ± 0.676 | 6.443 ± 0.575 | 6.784 ± 0.550 | 7.473 ± 0.953 |
| **7 dB** | eLORETA | 6.367 ± 0.247 | 7.197 ± 0.249 | 7.630 ± 0.218 | 9.923 ± 0.050 |
| | LCMV | 6.240 ± 0.107 | 6.239 ± 0.129 | 7.541 ± 0.112 | 9.899 ± 0.045 |
| | BC-VARETA | 5.117 ± 0.806 | 5.608 ± 0.726 | 5.750 ± 0.698 | 6.356 ± 1.176 |
| **0 dB** | eLORETA | 6.222 ± 0.149 | 7.117 ± 0.150 | 7.564 ± 0.136 | 9.865 ± 0.071 |
| | LCMV | 6.083 ± 0.069 | 5.780 ± 0.075 | 7.561 ± 0.080 | 9.909 ± 0.042 |
| | BC-VARETA | 4.488 ± 1.075 | 4.694 ± 1.107 | 4.783 ± 0.955 | 4.682 ± 1.249 |

| **Connectivity performance based on Cartesian Earth Movers' Distance (CEMD-USP)** | | | | | |
|---|---|---|---|---|---|
| | | **Connectivity type** | | | |
| **SNR** | Method | unconnected | randomly-connected (1<->3) | randomly-connected (1<->2) (1<->3) | fully-connected |
| **19 dB** | eLORETA | 524.2 ± 18.12 | 510.2 ± 20.71 | 524.4 ± 30.15 | 526.6 ± 32.67 |
| | LCMV | 485.8 ± 40.50 | 511.1 ± 42.66 | 545.9 ± 49.57 | 583.6 ± 48.32 |
| | BC-VARETA | 111.1 ± 10.53 | 135.0 ± 12.68 | 148.6 ± 13.02 | 203.0 ± 17.75 |
| **7 dB** | eLORETA | 562.2 ± 29.16 | 565.3 ± 25.84 | 567.1 ± 27.07 | 570.3 ± 37.39 |
| | LCMV | 567.7 ± 33.14 | 578.4 ± 38.83 | 568.1 ± 47.91 | 634.9 ± 37.39 |
| | BC-VARETA | 104.1 ± 12.06 | 129.2 ± 10.41 | 139.4 ± 14.04 | 201.5 ± 12.62 |
| **0 dB** | eLORETA | 580.9 ± 24.91 | 582.4 ± 23.22 | 566.3 ± 24.32 | 563.6 ± 32.70 |
| | LCMV | 571.4 ± 19.41 | 585.1 ± 44.77 | 566.7 ± 49.14 | 628.5 ± 40.15 |
| | BC-VARETA | 95.29 ± 9.452 | 122.1 ± 11.74 | 143.4 ± 12.84 | 192.4 ± 17.20 |

### 3.2 Real Data Analysis

#### 3.2.1 EEG Example

EEG data was selected from the Cuban Brain Mapping Project (*Hernandez-Gonzalez, 2011*), created in 2005 with the aim to obtain brain atlases of Cuban population. The EEG under analysis belongs to 32 years old healthy male in resting state (eyes closed) condition, recorded in 128 channels of MEDICID 5 system at sampling frequency of 200 Hz. The DEC was computed by extracting the Fourier coefficients samples from 600 time windows of 2 seconds each and the EEG SDTO (Lead Field) by the subject specific T1MRI, following the left branch (MEEG data processing) of the Toolbox, see Figure 6.

In Figure 10 we show the reconstructed source activity and connectivity with BC-VARETA at a single frequency component 10.57 Hz, belonging to the EEG alpha band (8 Hz – 13 Hz). BC-VARETA results are in congruence with previous studies on the physiology of human EEG. The neural correlates of resting state EEG in the spectral domain has been reported in previous works, in either eyes-closed or eyes-open conditions, unveiling the specific spatial signatures of different frequency bands (*Barry, 2007*). These findings and the results of our method suggest strong association of the Left/Right occipital lobe activity with alpha band. Our analysis reflected a dense occipital connectivity pattern, in the short-range (within Left and Right areas) and in the long-range (between Left and Right areas), something physiologically plausible as previous DTI studies have evidenced (*Dougherty et al., 2015*).

We performed further analysis by extracting the source activity and connectivity along the whole spectra, see Video 1 in link below. This results support what was discussed about the neural correlates of the MEEG alpha band, when sweeping the source activity and connectivity within the interval (8 Hz – 13 Hz). The solution for other spectral components, provided by this video, confirm previous studies, e.g. the generation of delta band (1.5 Hz – 3.5 Hz) by frontal lobe activity. A well-known fact, that is confirmed by BC-VARETA reconstruction along the whole spectra, is the smooth scrolling of source activity from frontal areas (lower frequencies 1.5Hz) across parietal and temporal lobe (mid frequencies) towards occipital areas (higher frequencies 13Hz).

***Video 1:*** *EEG processing*

#### 3.2.2 MEG Example

MEG data was selected from the first release of Human Connectome Project (*Marcus et.al., 2011*), large study of connectivity in healthy populations. The MEG under analysis belongs to a healthy male for eyes open resting state condition and attention fixation on a projected red crosshair. The data was recorded in 248 magnetometer channels of a MEG system at sampling frequency of 508 Hz. The DEC was computed by extracting the Fourier coefficients samples from 400 time windows of 5.08 seconds each and the MEG SDTO (Lead Field) by the subject specific T1MRI, following the left branch (MEEG data processing). In Figure 11 shows, in analogy to the EEG study, the reconstructed source activity and connectivity with BC-VARETA at 10.55 Hz and Video 2 below the whole spectra analysis. BC-VARETA results for MEG are in congruence with previous physiological studies and with those reported for the EEG, at both levels of analysis: sources activity and connectivity, as it is expected given the common neural substrate of these techniques.

***Video 2:*** *MEG processing*

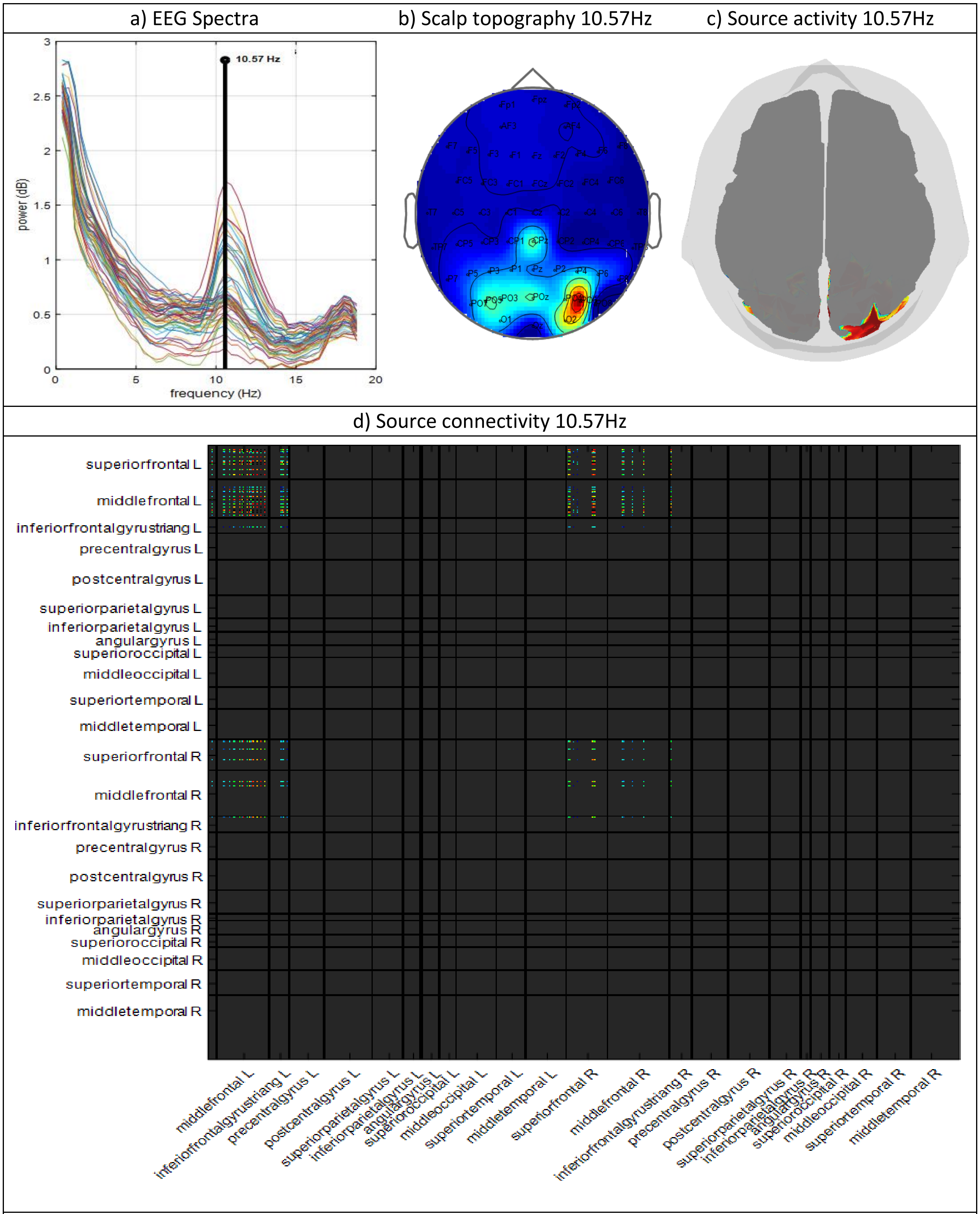

***Figure 10****. Real EEG data for a healthy subject in resting state (eyes closed) condition. BC-VARETA reconstruction was obtained at 10.57 Hz (typical alpha band component). a) Power spectral density for 58 EEG channels. b) Scalp topography of 58 EEG channels at the frequency component of analysis (10.57 Hz). c) Reconstructed source activity with BC-VARETA. d) Reconstructed source connectivity with BC-VARETA at a cortical subspace.*

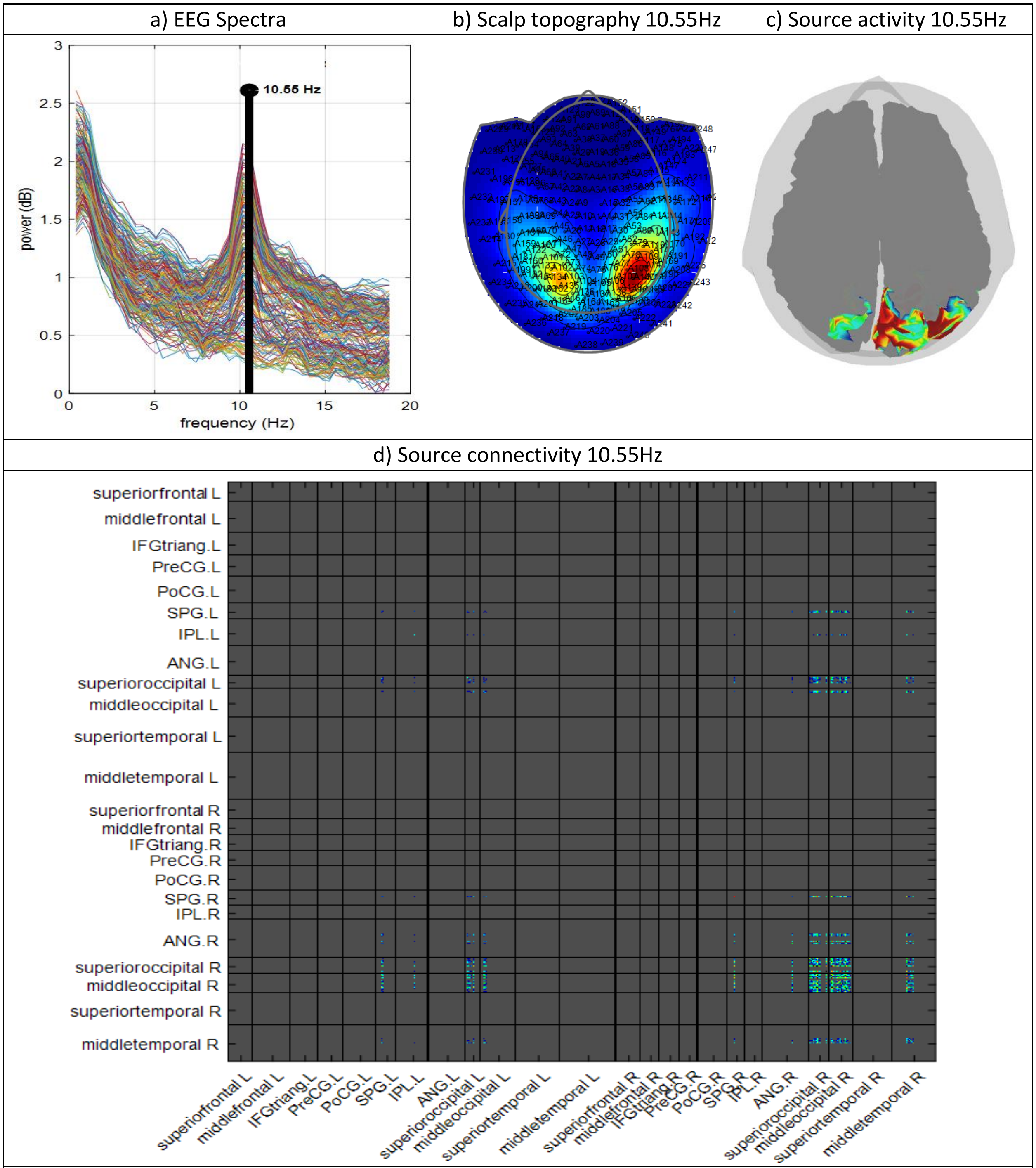

***Figure 11**. Real MEG data for a healthy subject in resting state (eyes open) condition, with attention fixation on a projected red crosshair. BC-VARETA reconstruction was obtained 10.55 Hz (typical alpha band component). a) Power spectral density for 248 MEEG magnetometer channels. b) Scalp topography of 248 EEG channels at the frequency component of analysis (10.55 Hz). c) Reconstructed source activity with BC-VARETA. d) Reconstructed source connectivity with BC-VARETA at a cortical subspace.*

## 4 CONCLUSIONS

The proposed tool demonstrated high performance in finding the neural signature (source activity and connectivity) of scalp MEEG signals. This can be ensured given the realistic simulation scenario (Simulation Benchmark) in which its ability for source analysis was tested, i.e. the multiple degrees of sparsity (super resolution) variability (different source configurations and connectivity modes) and realism (presence of noise in generators and sensors, and inverse crime evaluation). In complicated simulations BC-VARETA performance was better than well stablished methods, which operate under different assumptions, i.e. eLORETA and LCMV. These results were supported by sensitive quality measures (the typical and cartesian Earth Movers Distance), that were presented along with the well stablished ROC measures and M Box statistical test. Also, the results in real data (EEG, MEEG) revealed very interpretable neural correlates, when compared with previous studies.

We presented in detail the technical route of the whole methodology: model, inference, pseudocodes of both source activity and connectivity analysis and pseudocodes of simulations. This first release "BC-VARETA 1.0" constitutes a highly functional tool for either validation of methods and MEEG analysis, in multiple scenarios of source activity and connectivity.

## 5 ACKNOWLEDGEMENTS

This study was funded by the Grant No. 61673090 from the National Nature Science Foundation of China.